\documentclass[english]{IEEEtran}
\usepackage[T1]{fontenc}
\usepackage[latin9]{inputenc}
\usepackage{mathrsfs}
\usepackage{mathtools}
\usepackage{amsmath}
\usepackage{amssymb}
\usepackage{graphicx}

\makeatletter

\providecommand{\tabularnewline}{\\}

\usepackage{cuted}
\usepackage{flushend}

\@ifundefined{showcaptionsetup}{}{%
 \PassOptionsToPackage{caption=false}{subfig}}
\usepackage{subfig}
\makeatother

\usepackage{babel}
\begin{document}
\title{Can we imitate the principal investor's behavior to learn option price?}
\author{Xin Jin\thanks{X. Jin is with Bank of Montreal, Canada. (e-mail: Xin.Jin@bmo.com;
felixxinjin@gmail.com).}\\
Bank of Montreal}
\maketitle
\begin{abstract}
This paper presents a framework of imitating the principal investor's
behavior for optimal pricing and hedging options. We construct a non-deterministic
Markov decision process for modeling stock price change driven by
the principal investor's decision making. However, low signal-to-noise
ratio and instability that are inherent in equity markets pose challenges
to determine the state transition (stock price change) after executing
an action (the principal investor's decision) as well as decide an
action based on current state (spot price). In order to conquer these
challenges, we resort to a Bayesian deep neural network for computing
the predictive distribution of the state transition led by an action.
Additionally, instead of exploring a state-action relationship to
formulate a policy, we seek for an episode based visible-hidden state-action
relationship to probabilistically imitate the principal investor's
successive decision making. Unlike conventional option pricing that
employs analytical stochastic processes or utilizes time series analysis
to model and sample underlying stock price movements, our algorithm
simulates stock price paths by imitating the principal investor's
behavior which requires no preset probability distribution and fewer
predetermined parameters. Eventually the optimal option price is learned
by reinforcement learning to maximize the cumulative risk-adjusted
return of a dynamically hedged portfolio over simulated price paths.
\end{abstract}


\section{Introduction}

Modern quantitative finance employs the key idea of replicating a
portfolio to bind financial derivative pricing to cross-sectional
variation in underlying stock returns. Therefore, an effective way
to model and simulate the distribution of future stock price movements
is in demand and pivotal. Traditional models attempt to mirror the
stock price movements as an analytical stochastic process or a combination
of multiple random processes, such as the geometric Brownian motion
in the Black--Scholes model \cite{BS model}, the compound Poisson
process in the Merton jump diffusion model \cite{MertonJumpDiffusionModel},
the mean reverting square-root process combining two Wiener processes
in Heston model \cite{HestonModel}, skewed Student's t-distribution
for the standardized residuals in $GARCH\left(p,q\right)$ \cite{GARCH}.
However, framing the evolution of stock prices into stochastic processes
is either too basic to cope with complexity or it relies on numerous
undetermined parameters to be calibrated frequently. 

This downside motivates us to develop a data-driven way to simulate
underlying price movements to dispense with calibrating parameters
by imitating the principal investor's behavior and dynamically hedge
portfolio over simulated price paths to learn option price. The Principal
Investor (PI) refers to the collection of investors who have a leading
influence on the stock market and the PI's decision indicates the
quantity of the capital that the PI chooses to invest in or sell out
a particular underlying stock in the given time slot. We build our
algorithm in three steps: 1. compute the predictive distribution of
stock price change led by the PI's decision, 2. imitate the PI\textquoteright s
successive decision making (the PI's behavior) to simulate stock price
paths, 3. learn option price to maximize the cumulative risk-adjusted
return of a dynamically hedged portfolio over simulated price paths.

The interaction between the PI's decision and stock price changes
can be described by a non-deterministic Markov Decision Process (MDP)
\cite{MDP}. At each time step, MDP is in a state (spot price of the
underlying stock) and the decision maker chooses an action (the PI's
decision) based on the current state. MDP then randomly moves into
a new state with a probability that is influenced by the chosen action. 

The predictive distribution of stock price change led by the PI's
decision instead of a fixed nonlinear relationship between stock price
change and the PI's decision is desired due to the reason that MDP
is non-deterministic. For the purpose of deriving this predictive
distribution from data, we resort to Bayesian Deep Neural Network
(B-DNN) \cite{BNN-1,BNN-2,BNN-3,BNN-4} which introduces Bayesian
statistics to Deep Neural Network (DNN) weights inference. Against
a point estimate of DNN weights, B-DNN learns a probability distribution
over the network weights. B-DNN predicts the probability distribution
over price change induced by the PI's decision through a two-stage
process (1. Bayesian inference, 2. regression analysis). In the first
stage, the distribution over B-DNN weights is inferred from the training
dataset. Whereafter, a regression analysis is performed to model price
change as a posterior predictive distribution which depends on the
posterior distribution of DNN weights obtained from the first stage.
Owing to the merit of DNN, the second step does away with a closed-form
expression for non-linear multi-dimensional feature mapping. Instead,
the regression analysis is built on a linear combination of basis
vectors with few parameters to be pre-set. The numeric results based
on real stock market data reveal that the PI's decision and the stock
price change are approximately positively correlated. We apply leader-follower
type Keynesian beauty contest \cite{Keynes1,Keynes2} to explain the
approximately positive correlation between the stock price change
and the PI's decision: a single retail investor easily believes that
the instantaneous observation of the PI\textquoteright s decision
is the most investors\textquoteright{} concurrent decision to follow. 

Leader-follower type Keynesian beauty contest discloses the behavior
pattern of a single retail investor but not the PI's strategy to make
a decision that is also the last remaining piece of our MDP. In contrast
to conventional MDP that associates a single type of factor (i.e.,
state) to the action selection by a policy function, we believe that
the action selection depends on two types of factors (i.e., visible
state and hidden state) and build a Visible-Hidden Markov Network
to model the dependency of the PI\textquoteright s decision making
on visible factor and hidden factor. An algorithm is proposed to update
the visible-hidden Markov network for best possibly fitting the visible
state sequence and the observation sequence. Through the instrumentality
of a trained visible-hidden Markov network for imitating the PI's
behavior from observations, we simulate the paths of the PI\textquoteright s
successive decision making and further transform them to stock price
paths through Bayesian inference and regression analysis.

The final step to achieve our ultimate goal is designing an optimization
criterion for pricing the option and a dynamic hedging algorithm that
can efficiently take advantage of the cross-sectional information
yielded by simulated price paths of the underlying stock. To this
end, a self-financing portfolio is built to dynamically replicate
the option price. The self-financing restriction raises a risk of
exhausting cash for rebalancing the portfolio during the life of the
option owing to a volatile price of the underlying stock. With the
intention of minimizing this risk, the option price consists of a
risk-free price and a risk-adjusted cost. The self-financing restriction
also guarantees that the portfolio's value change between adjacent
time steps comes from the portfolio's return. Thus, we are capable
of exploiting reinforcement learning to maximize the cumulative risk-adjusted
return for recursively learning the hedging position and pricing the
option.

\section{Related Work}

In the financial market, traditional Time Series Analysis (TSA) approaches
play a dominant role to generate distribution for future stock prices.
The linear TSA model, for example, Autoregressive Moving Average ($ARMA\left(p,q\right)$)
forecasts the price through a linear regression that relates the targeted
value to a linear combination of $p$ previous lag values, current
residual error and $q$ past residual errors. $ARMA\left(p,q\right)$
is reliable only if the time series is stationary (in the sense of
mean, variance, and covariance) without seasonality. $ARMA\left(p,q\right)$
is extended to cover the situation that the time series exhibits a
trend and seasonal variation. ARMA Autoregressive Integrated Moving
Average ($ARIMA\left(p,d,q\right)$) \cite{arima} introduces differencing
to $ARMA\left(p,q\right)$ to eliminate the non-stationarity in the
sense of mean (i.e. the trend). Seasonal ARIMA $\left(ARIMA\left(p,d,q\right)\times\left(P,D,Q\right)S\right)$
\cite{seasonal arima} incorporates additional seasonal differencing
AR and MA terms into the non-seasonal ARIMA to deal with seasonal
variation. The non-linear TSA models (Autoregressive Conditional Heteroskedasticity
($ARCH\left(p\right)$) / Generalized Autoregressive Conditional Heteroskedasticity
($GARCH\left(p,q\right)$) models) are employed to forecast the time
series which is non-stationary in the sense of variance (time-varying
volatility). $ARCH\left(p\right)$ models the variance at a time step
as a linear combination of $p$ previous lag squared residual errors.
$GARCH\left(p,q\right)$ model adds another linear combination of
$q$ previous lag variance to $ARCH\left(p\right)$. The non-linear
models have been improved in recent studies, for example, \cite{ARCH skewness}
works on fitting skewness to make distribution assumptions about the
residuals as financial time series data often has a deviation from
a normal distribution. In practice, the linear TSA model and the non-linear
TSA model are commixed to generate the distribution for future stock
prices. For instance, $ARMA\left(p,q\right)$ is used to forecast
the mean and $GARCH\left(p,q\right)$ gives the volatility prediction.

Recently deep neural network-based approaches for stock price prediction
have attracted much attention. \cite{Fischer,Recurrent neural network}
propose to apply recurrent neural networks, such as Long short-term
memory (LSTM) networks and gated recurrent unit (GRU) to financial
time series predictions. Graph Neural Networks (GNN) are considered
as an effective deep learning model for stock movement prediction
in \cite{Graph1,Graph2}. In \cite{CNN1,CNN2}, authors recommend
Convolutional Neural Network (CNN) as a novel method for predicting
stock price movements.

Our proposed approach generates the distribution of stock price by
imitating the principal investor's behavior based on B-DNN and visible-hidden
Markov. Compared with TSA approaches, our approach does not require
many pre-determined parameters and is not reliant on any preset probability
distribution. As a counter example, we have to decide order $\left(p,q\right)$
through statistical tests and specify the distribution assumptions
about the residuals before using $GARCH\left(p,q\right)$. In addition,
our proposed approach generates the distribution of stock price instead
of predicting the stock price as above listed recent deep neural network-based
approaches are used for.

\section{Effect of the principal investor's decision\label{sec:Effect-of-the}}

As described in the introduction section, we are motivated to model
the evolution of the underlying stock price by virtue of DNN and behavioral
analysis, rather than analytically characterize stochastic processes
in a traditional way. With this end in view, a non-deterministic MDP
is designed to associate stock price evolution with the PI's successive
decision making. This section is on a quest for the impact of PI's
decision on stock price change.

We define PI as the aggregation of investors who are considered as
the dominating force in the equity market. PI could be composed of
large institutional investors or a sizable group of retail investors
acting in concert. PI's decision is concerned with the amount of money
flowing into or out of a stock within a time slot. We approximately
quantize PI's decision as a signed number $d\in\varOmega_{d}$ in
terms of the opening price $o\in\varOmega_{o}$, the highest price
$h$, the lowest price $l$, the closing price $c\in\varOmega_{c}$,
and the volume $v$:

\begin{equation}
d=\left(vh\frac{h-o}{h-l}\right)\mathrm{\textrm{sign}\left(\mathit{c-o}\right)}.\label{eq:decision}
\end{equation}
This representation allows the difference between the opening and
closing prices to indicate the direction of the money flow. Moreover,
the difference between the opening and highest prices becomes a primary
determinant of the money flow volume. Through simple math deduction,
this synthetic variable can be proved to have a monetary unit, and
we consider it as USD in this paper.

The sample spaces $\varOmega_{d}$ and $\varOmega_{o}$ (or $\varOmega_{c}$)
constitute respectively the MDP action space and state space. The
price change functions as an indicator of the MDP state transition:
\begin{equation}
g=\frac{c}{o}-1\label{eq:price change}
\end{equation}
We presume that the price change is incompletely impacted by PI's
decision and partly random. This implies that the price change follows
a probability distribution which is dependent on PI's decision. We
intend to use a data-driven regression method to determine this probability
distribution and verify our presumption. For this aim, we resort to
B-DNN \cite{BNN-1,BNN-2,BNN-3,BNN-4} which applies Bayesian statistics
to DNN weights inference. By further merging B-DNN with Gaussian process
\cite{DNN2GP,GP}, the price change distribution is simplified as
a gaussian distribution.

Given a dataset $\mathcal{\mathscr{D}}\coloneqq\left\{ \left(d_{n},g_{n}\right)\right\} _{n=0}^{\mathrm{T}-1}$which
contains pairs of quantized PI's decision and the price change sampled
from $\mathrm{T}$ consecutive time slots, we apply gradient ascent
coupled with the dataset $\mathscr{D}$ to train a DNN. The DNN takes
an input data $d_{n}$ and returns a predicted value $f_{\boldsymbol{\theta}}\left(d_{n}\right)\in\mathbb{R}$
to compare with the observed value $g_{n}$ for producing the empirical
loss, where $\boldsymbol{\theta=}\left[\theta_{1},\ldots,\theta_{M}\right]^{T}$
signifies the DNN weights. The loss is defined in terms of the loss
function $l\left(g_{n},f_{\boldsymbol{\theta}}\left(d_{n}\right)\right)$
and the $L2$ regularizer $\mathscr{\mathit{r}}\left(\boldsymbol{\theta}\right)$
with a regularization parameter $\lambda$: 
\begin{align}
\mathscr{L}\left(\boldsymbol{\theta},\mathscr{D}\right) & =L\left(\boldsymbol{\theta},\mathscr{D}\right)+\mathscr{\mathit{r}}\left(\boldsymbol{\theta}\right)\nonumber \\
 & =\sum_{n=0}^{\mathrm{T}-1}l\left(g_{n},f_{\boldsymbol{\theta}}\left(d_{n}\right)\right)+\frac{\lambda}{2}\boldsymbol{\theta}^{T}\boldsymbol{\theta}.\label{eq:loss}
\end{align}

In contrast to fit a point estimate of DNN weights, B-DNN learns a
distribution over the weights. We take advantage of this merit of
B-DNN to infer the posterior distribution over the weights (Bayesian
inference) and therefore derive a posterior predictive distribution
which measures the distribution over price change induced by PI's
decision (Regression analysis). 

In order to implement Bayesian inference on the weights, the loss
is straightforwardly embedded in the posterior distribution over the
weights via the exponential family form $P\left(\boldsymbol{\theta}\mid\boldsymbol{\mathscr{D}}\right)\coloneqq\frac{e^{\mathscr{-L}\left(\boldsymbol{\theta},\mathscr{D}\right)}}{\int e^{\mathscr{-L}\left(\boldsymbol{\theta},\mathscr{D}\right)}d\boldsymbol{\theta}}$.
It implies both the likelihood and the prior also belong to the exponential
family with the expression as $P\left(\mathscr{D}\mid\boldsymbol{\theta}\right)\coloneqq\frac{e^{-L\left(\boldsymbol{\theta},\mathscr{D}\right)}}{\int e^{-L\left(\boldsymbol{\theta},\mathscr{D}\right)}d\boldsymbol{\theta}}$
and $P\left(\boldsymbol{\theta}\right)\coloneqq\frac{e^{\mathit{-r}\left(\boldsymbol{\theta},\mathscr{D}\right)}}{\int e^{\mathit{-r}\left(\boldsymbol{\theta},\mathscr{D}\right)}d\boldsymbol{\theta}}$.
In such a manner, the Maximum A Posteriori (MAP) estimation $\widehat{\boldsymbol{\theta}}=\underset{\boldsymbol{\theta}}{\arg\max}\,P\left(\boldsymbol{\theta}\mid\boldsymbol{\mathscr{D}}\right)$
is equivalent to minimize mean square error. However, the immediate
MAP estimation is often computationally impractical. 

Laplace approximation \cite{LaplaceApproximation0,LaplaceApproximation1,LaplaceApproximation2}
provides a computationally feasible Bayesian inference on weights.
The main idea is to approximate likelihood times prior with a Gaussian
density function in respect that likelihood times prior is a well-behaved
uni-modal function in $\mathcal{L}^{2}$ space. The posterior distribution
over the weights is consequently derived to be the below Gaussian
distribution (see derivation in Appendix A):
\begin{equation}
P\left(\mathrm{\boldsymbol{\theta}}|\mathcal{\mathscr{D}}\right)\approx\mathcal{N}\left(\boldsymbol{\theta}\mid\boldsymbol{\theta}^{*},\left[\nabla_{\boldsymbol{\theta\theta}}^{2}L\left(\boldsymbol{\theta}^{*},\mathscr{D}\right)+\lambda\mathbf{I}_{\mathit{M}}\right]^{-1}\right),\label{eq:posterior}
\end{equation}
where $\boldsymbol{\theta}^{*}=\left[\theta_{1}^{*},\ldots,\theta_{M}^{*}\right]^{T}$
is the mode of $P\left(\mathscr{D}\mid\boldsymbol{\theta}\right)P\left(\boldsymbol{\theta}\right)$.

After obtaining the posterior over the weights, we then seek for the
posterior predictive distribution over price change $P\left(g^{*}\mid d^{*},\mathscr{D}\right)$,
where $d^{*}\in\mathbb{R}$ denotes a test sample of quantized PI's
decision and $g^{*}$ signifies the corresponding observation of the
price change. The high dimensional projection and the nonlinear relationship
between $d^{*}$ and $f_{\boldsymbol{\theta}^{*}}\left(d^{*}\right)$
impede to directly find $P\left(g^{*}\mid d^{*},\mathscr{D}\right)$
in closed form. In order to overcome this impediment, a new random
variable $y$ which has a readily accessible posterior predictive
distribution is constructed and can be linearly expressed in terms
of $g^{*}$. As a result, $P\left(g^{*}\mid d^{*},\mathscr{D}\right)$
is inferred from the given posterior predictive distribution $P\left(y\mid d^{*},\mathscr{D}\right)$.
As presented in \cite{DNN2GP}, $y$ is possible to be designed as
a linear combination of a Gaussian process and additive gaussian noise.
Different from \cite{DNN2GP}, we calculate the parameters of the
Gaussian process (i.e., $m\left(d^{*}\right)$ in Equation (\ref{eq:mean function})
and $k\left(d^{*},d^{*}\right)$ in Equation (\ref{eq:covariance function}))
based on $P\left(\mathrm{\boldsymbol{\theta}}|\mathcal{\mathscr{D}}\right)$
instead of $P\left(\mathrm{\boldsymbol{\theta}}|\mathcal{\mathscr{\mathscr{D}^{\prime}}}\right)$
where $\mathscr{D}^{\prime}\coloneqq\left\{ \left(y_{n},c_{n}\right)\right\} _{n=0}^{\mathrm{T}-1}$
is a transformed dataset and $y_{n}$ is a sample of the constructed
variable. In addition, we consider the residual ($\varepsilon$ in
Equation (\ref{eq: constructed variable with square loss})) as additive
independent identically distributed (i.i.d) Gaussian noise rather
than additionally bring in the second-order partial derivatives of
the loss to represent noise. 

The random variable $y$ is designed as the following combination
of the standard linear regression model and the first order derivative
of the loss function:

\begin{equation}
y=\boldsymbol{\phi}\left(d^{*}\right)^{T}\boldsymbol{\theta}+\nabla_{f_{\boldsymbol{\theta}^{*}}\left(d^{*}\right)}l\left(g^{*},f_{\boldsymbol{\theta}^{*}}\left(d^{*}\right)\right),\label{eq: constructed variable}
\end{equation}
where DNN weights follow a prior distribution $\boldsymbol{\theta}\sim\mathcal{N}\left(\mathbf{0},\mathbf{K}_{\boldsymbol{\theta\theta}}\right)$
with the covariance matrix $\mathbf{K}_{\boldsymbol{\theta\theta}}=\lambda^{-1}\mathbf{I}_{\mathit{M}}$.
And $\boldsymbol{\phi}\left(d^{*}\right)$ is a Jacobian matrix whose
entries form a set of basis functions to project the input sample
into $M$ dimensional feature space: 

\begin{align}
\boldsymbol{\phi}\left(d^{*}\right) & =\nabla_{\boldsymbol{\theta}}f_{\boldsymbol{\theta}}\left(d^{*}\right)\mid_{\boldsymbol{\theta}=\boldsymbol{\theta}^{*}}\nonumber \\
 & =\left[\begin{array}{c}
\frac{\partial f_{\boldsymbol{\theta}}\left(d^{*}\right)}{\partial\theta_{1}}\mid{}_{\theta=\theta_{1}^{*}}\\
\vdots\\
\frac{\partial f_{\boldsymbol{\theta}}\left(d^{*}\right)}{\partial\theta_{M}}\mid_{\theta_{M}=\theta_{M}^{*}}
\end{array}\right],\label{eq:Jacobian matrix}
\end{align}
As the linear regression model is built on the feature space in Equation
(\ref{eq:Jacobian matrix}), it successfully leaves the difficulty
of processing non-linear high-dimensional mapping back to DNN.

\begin{figure}
\subfloat[\label{fig:1a}]{\includegraphics[width=4.5cm,height=3.5cm]{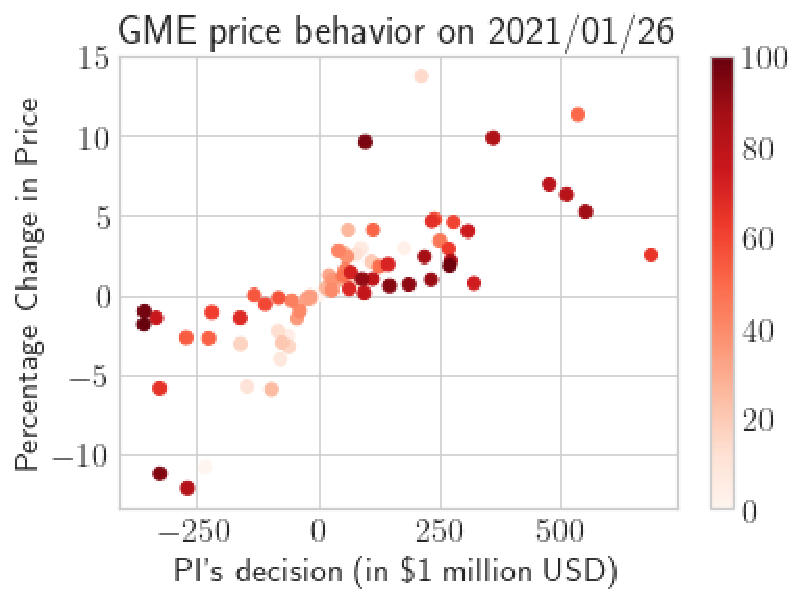}}\subfloat[\label{fig:1b}]{\includegraphics[width=4.5cm,height=3.5cm]{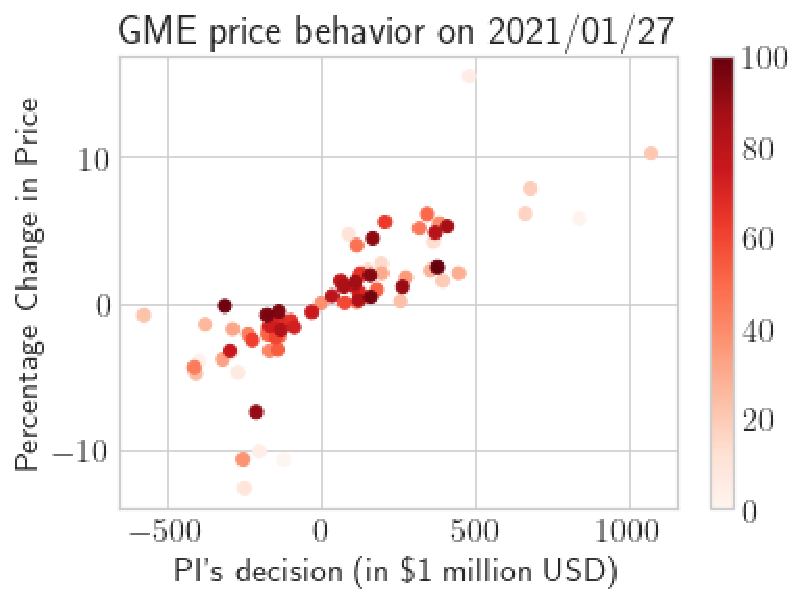}}\medskip{}

\subfloat[\label{fig:1c}]{\includegraphics[width=4.5cm,height=3.5cm]{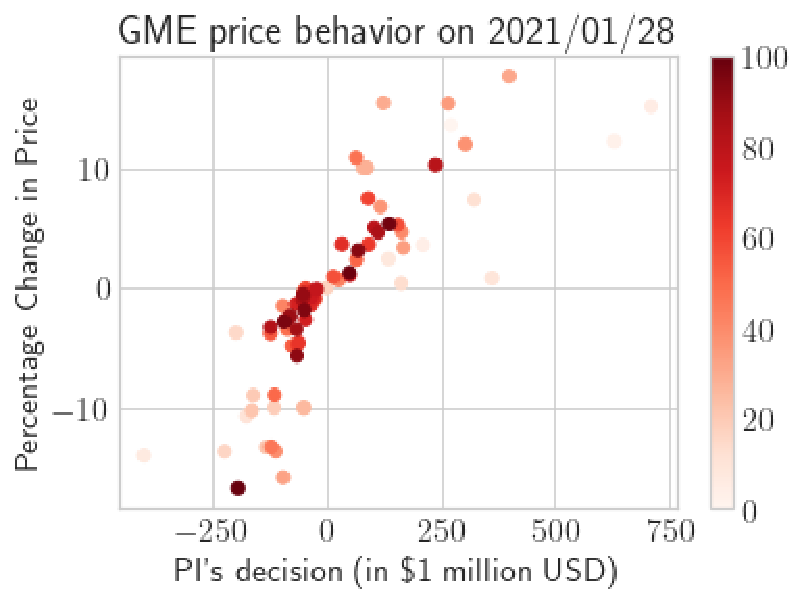}}\subfloat[\label{fig:1d}]{\includegraphics[width=4.5cm,height=3.5cm]{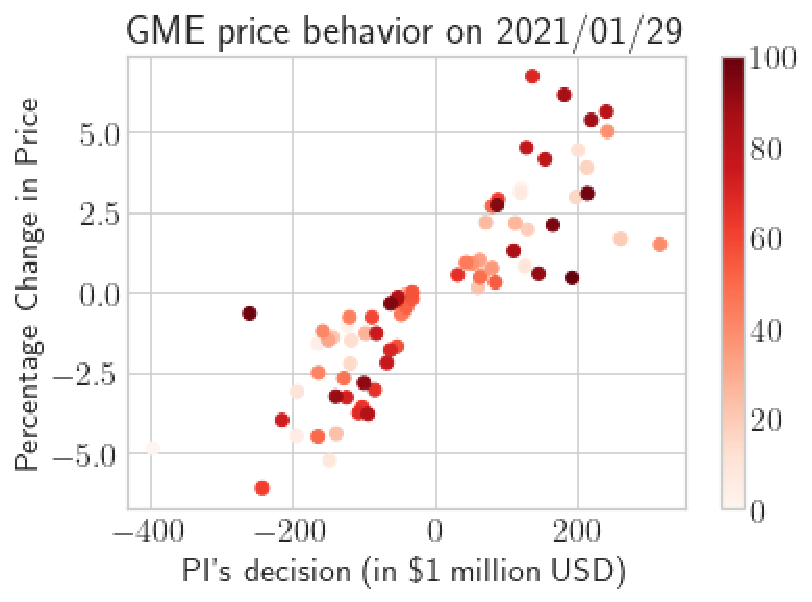}}\medskip{}

\subfloat[\label{fig:1e}]{\includegraphics[width=4.5cm,height=3.5cm]{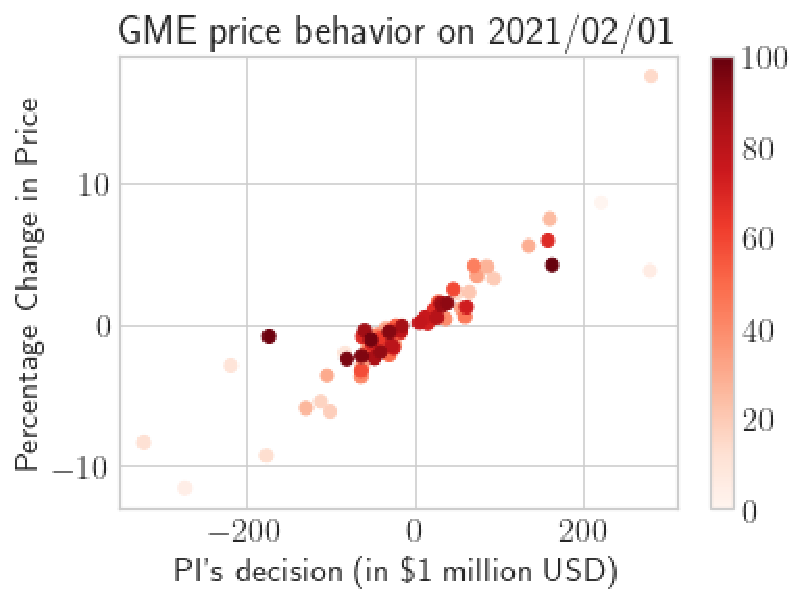}}\subfloat[\label{fig:1f}]{\includegraphics[width=4.5cm,height=3.5cm]{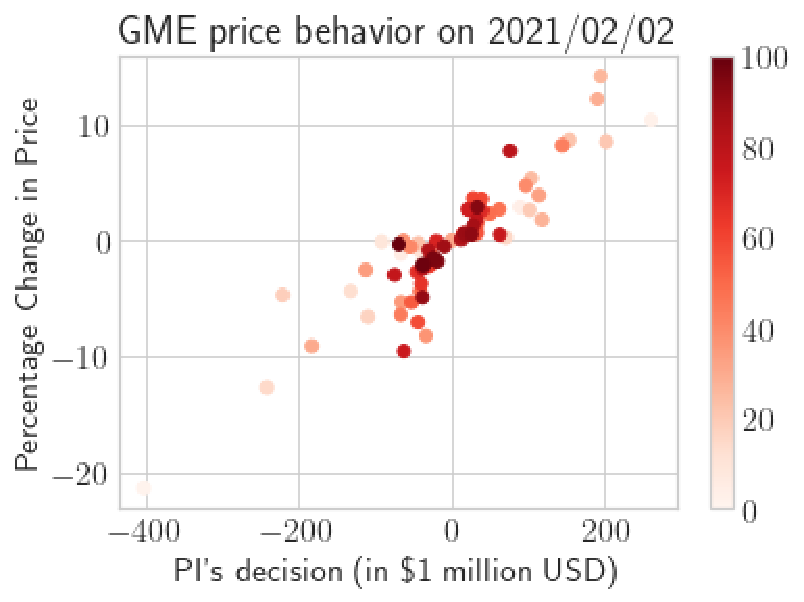}

}

\caption{GME behavior observed in each 5-minute time slot from 09:35 to 16:00
during 6 consecutive trading days. Earlier time slots are with lighter
shades.\label{fig:GME-behavior-observed}}
\end{figure}

For a quadratic loss function $l\left(g^{*},f_{\boldsymbol{\theta}^{*}}\left(d^{*}\right)\right)=\frac{1}{2}\left(g^{*}-f_{\boldsymbol{\theta}^{*}}\left(d^{*}\right)\right)^{2}$,
the constructed random variable $y$ can be simplified in terms of
$g^{*}$ from Equation (\ref{eq: constructed variable}):

\begin{align}
y & =\boldsymbol{\phi}\left(d^{*}\right)^{T}\boldsymbol{\theta}+\left(f_{\boldsymbol{\theta}^{*}}\left(d^{*}\right)-g^{*}\right)\nonumber \\
 & =\varphi\left(d^{*}\right)+\varepsilon,\label{eq: constructed variable with square loss}
\end{align}
where $f_{\boldsymbol{\theta}^{*}}\left(d^{*}\right)-g^{*}$ is the
residual that can be modeled as additive independent identically distributed
Gaussian noise $\varepsilon\sim\mathcal{N}\left(0,\sigma^{2}\right)$.
Gaussian process is an effective tool for describing a distribution
over functions \cite{GP}. From the function-space view, the process
$\varphi\left(d^{*}\right)$ is specified by a Gaussian process $\varphi\left(d^{*}\right)\sim\mathcal{GP}\left(m\left(d^{*}\right),k\left(d^{*},d^{*}\right)\right)$,
where we calculate the mean function $m\left(d^{*}\right)$ and the
covariance function $k\left(d^{*},d^{*}\right)$ based on the posterior
distribution over the weights $P\left(\mathrm{\boldsymbol{\theta}}|\mathcal{\mathscr{D}}\right)$
obtained in Equation (\ref{eq:posterior}):

\begin{align}
m\left(d^{*}\right) & =\boldsymbol{\phi}\left(d^{*}\right)^{T}\mathbb{E}\left[\boldsymbol{\theta}\mid\mathscr{D}\right]\nonumber \\
 & =\boldsymbol{\phi}\left(d^{*}\right)^{T}\boldsymbol{\theta}^{*},\label{eq:mean function}
\end{align}

\begin{align}
k\left(d^{*},d^{*}\right) & =\phi\left(d^{*}\right)^{T}\mathbb{E}\left[\boldsymbol{\theta\theta}^{T}\mid\mathscr{D}\right]\phi\left(d^{*}\right)\nonumber \\
 & =\phi\left(d^{*}\right)^{T}\left[\nabla_{\boldsymbol{\theta\theta}}^{2}L\left(\boldsymbol{\theta}^{*},\mathscr{D}\right)+\lambda\mathbf{I}_{\mathit{M}}\right]^{-1}\phi\left(d^{*}\right).\label{eq:covariance function}
\end{align}

Equation (\ref{eq: constructed variable with square loss}) exhibits
that the random process $y$ is a linear combination of a Gaussian
process and additive Gaussian noise. Therefore, the posterior predictive
distribution of $y$ can be easily deduced from the statistical description
of the Gaussian process and Gaussian noise as:

\begin{equation}
P(y\mid d^{*},\mathscr{D})=\mathcal{N}\left(\boldsymbol{\phi}\left(d^{*}\right)^{T}\boldsymbol{\theta}^{*},k\left(d^{*},d^{*}\right)+\sigma^{2}\right).\label{eq:post pred distri of y}
\end{equation}

Since we always prefer to perform a prediction based on the DNN weights
at the mode of the posterior $\boldsymbol{\theta}^{*}$ rather than
any possible DNN weights, only the determinate realization $\boldsymbol{\phi}\left(d^{*}\right)^{T}\boldsymbol{\theta}^{*}$
should be extracted from the Gaussian process instead of the entire
Gaussian process $\boldsymbol{\phi}\left(d^{*}\right)^{T}\boldsymbol{\theta}$
for calculating $g^{*}$. Thuswise $g^{*}$ is a linear combination
of a random variable and a constant: 

\begin{align}
g^{*} & =y-\boldsymbol{\phi}\left(d^{*}\right)^{T}\boldsymbol{\theta}^{*}+f_{\boldsymbol{\theta}^{*}}\left(d^{*}\right)\nonumber \\
 & =y+C.\label{eq:g =00003D process + constant}
\end{align}

Consequently, the posterior predictive distribution over price change
is derived from Equation (\ref{eq:post pred distri of y}) and (\ref{eq:g =00003D process + constant}): 

\begin{equation}
P(g^{*}\mid d^{*},\mathscr{D})=\mathcal{N}\left(f_{\boldsymbol{\theta}^{*}}\left(d^{*}\right),k\left(d^{*},d^{*}\right)+\sigma^{2}\right).\label{eq:predictive distri of g*}
\end{equation}

The observation of the price change $g^{*}$ shares the same posterior
predictive variance with $y$ owing to the fact that the predicted
value $\boldsymbol{\phi}\left(d^{*}\right)^{T}\boldsymbol{\theta}^{*}$
is a determinate realization produced at the mode of the posterior
$\boldsymbol{\theta}^{*}$ and the deduction of the predicted value
from $y$ is incapable of eliminating the immanent uncertainty in
the inference on weights and the residual. When we consider an input
vector of test samples $\mathbf{d}^{*}$=$\left[d_{1}^{*},\ldots,d_{N}^{*}\right]^{T}$,
Equation (\ref{eq:predictive distri of g*}) turns into the below
form:

\begin{equation}
P(g^{*}\mid\mathbf{d}^{*},\mathscr{D})=\mathcal{N}\left(f_{\boldsymbol{\theta}^{*}}\left(\mathbf{d}^{*}\right),diag\left(\mathit{k}\left(\mathbf{d}^{*},\mathbf{d}^{*}\right)+\sigma^{2}\mathbf{I}_{\mathit{N}}\right)\right).\label{eq:predictive distri of g* input vector}
\end{equation}

During Jan. 26, 2021-Feb. 02, 2021, GameStop Corporation (GME) stock
price swiftly rose and then fell with the unusually high price and
volatility because a short squeeze was triggered by organized retail
investors. We would like to explore the behavioral finance behind
this event through our model. The intraday data for the 6 consecutive
trading days is exercised to infer B-DNN coefficients and be referred
to produce testing samples for validating regression analysis. Due
to the unreachability, the pre/post-market data is not included even
though some huge price fluctuations occurred in the pre/post-market
sessions. We structure a dataset $\mathcal{D}$ as 77 samples in each
of 6 episodes to accommodate the equity pricing data in the period
from 09:35 to 16:00 for the 6 consecutive trading days. Each sample
is a five-dimensional vector which enumerates open-high-low-close
prices and volume observed in a 5-minute time slot. We discard the
pricing data in the first time slot (i.e., 09:30 to 09:35) because
the open price is subject to the pre-market trading and gaps up or
down from the previous day\textquoteright s close. Moreover, the volume
in the first time slot is extremely high and influenced by the pre-market
news. Based on Equation (\ref{eq:decision}), (\ref{eq:price change})
and dataset $\mathcal{D}$, we yield the training dataset $\mathcal{\mathscr{D}}_{i}\coloneqq\left\{ \left(d_{n},g_{n}\right)\right\} _{n=0}^{\mathrm{T}_{i}-1}$
which contains the pairs of quantized PI's decision and price change
sampled from $\mathrm{T}_{i}$ consecutive time slots in the $i$th
episode as illustrated in Figure \ref{fig:GME-behavior-observed}.
The samples are plotted in chronological order. The lighter shade
indicates that the sample is taken in an earlier time slot.

\begin{figure}
\subfloat[\label{fig:2a}]{\includegraphics[width=4.5cm,height=2.5cm]{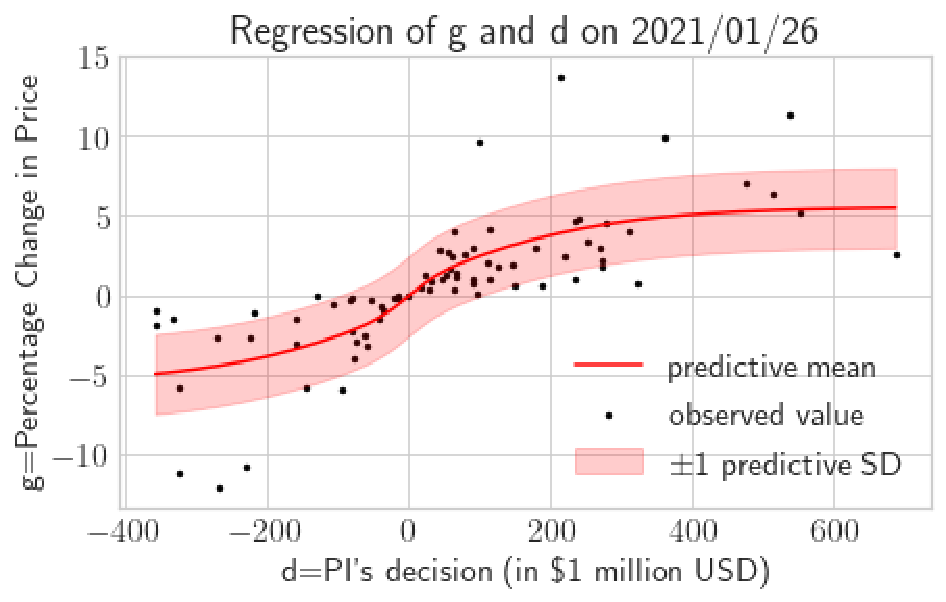}}\subfloat[\label{fig:2b}]{\includegraphics[width=4.5cm,height=2.5cm]{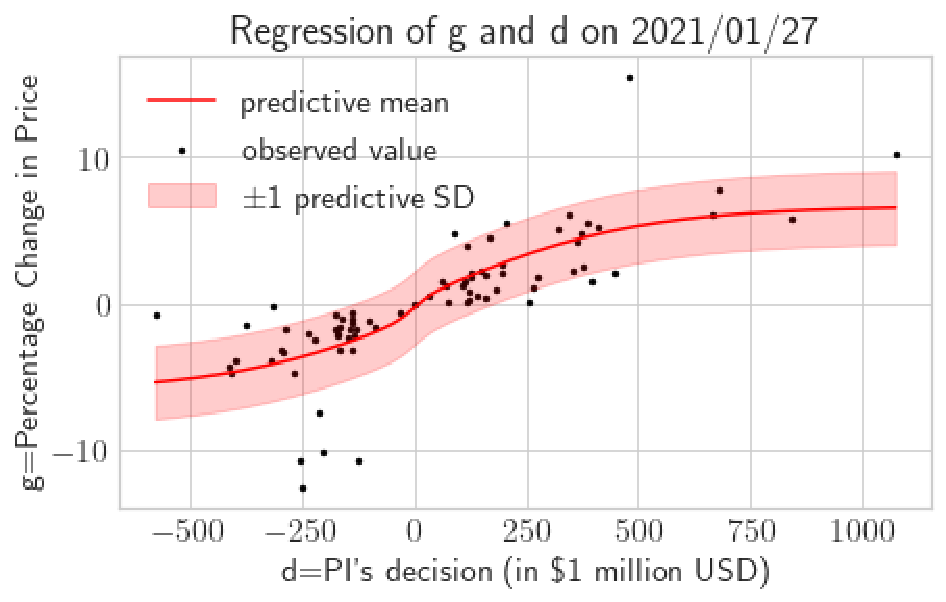}}\medskip{}

\subfloat[\label{fig:2c}]{\includegraphics[width=4.5cm,height=2.5cm]{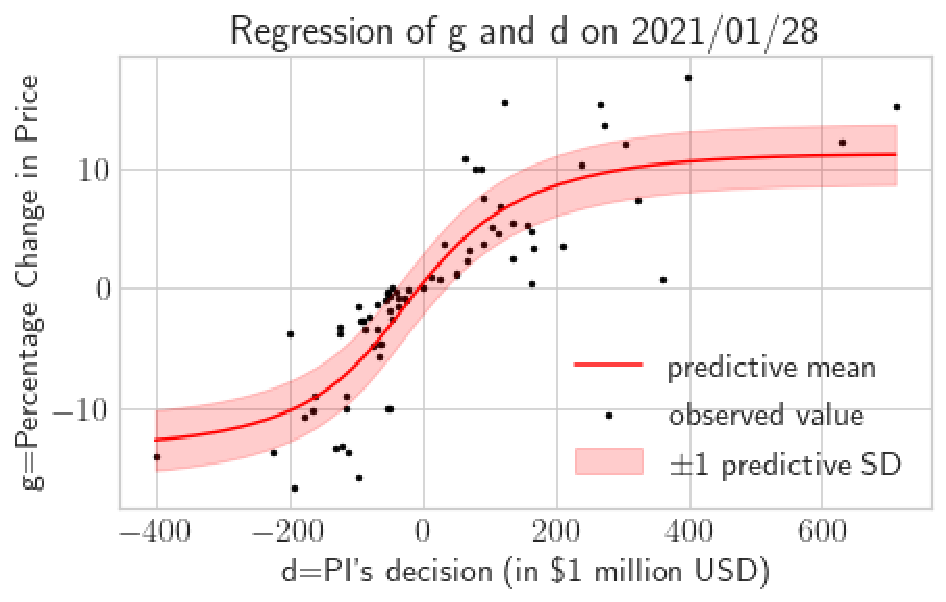}}\subfloat[\label{fig:2d}]{\includegraphics[width=4.5cm,height=2.5cm]{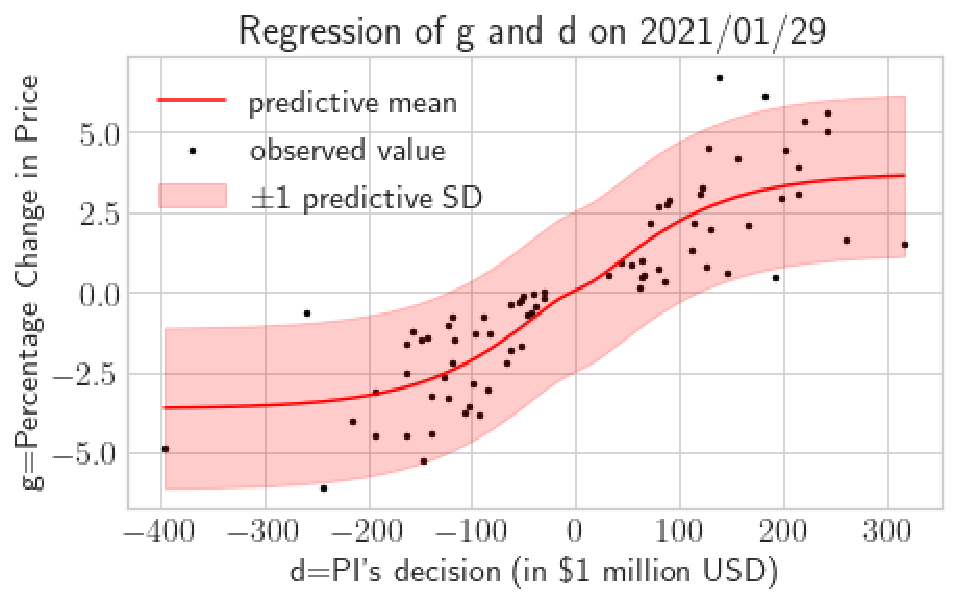}}\medskip{}

\subfloat[\label{fig:2e}]{\includegraphics[width=4.5cm,height=2.5cm]{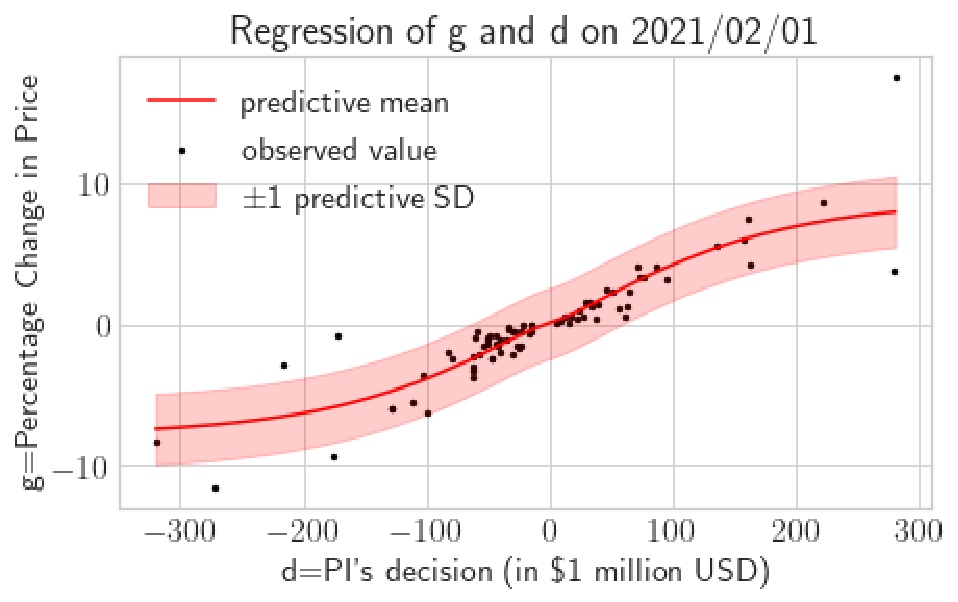}}\subfloat[\label{fig:2f}]{\includegraphics[width=4.5cm,height=2.5cm]{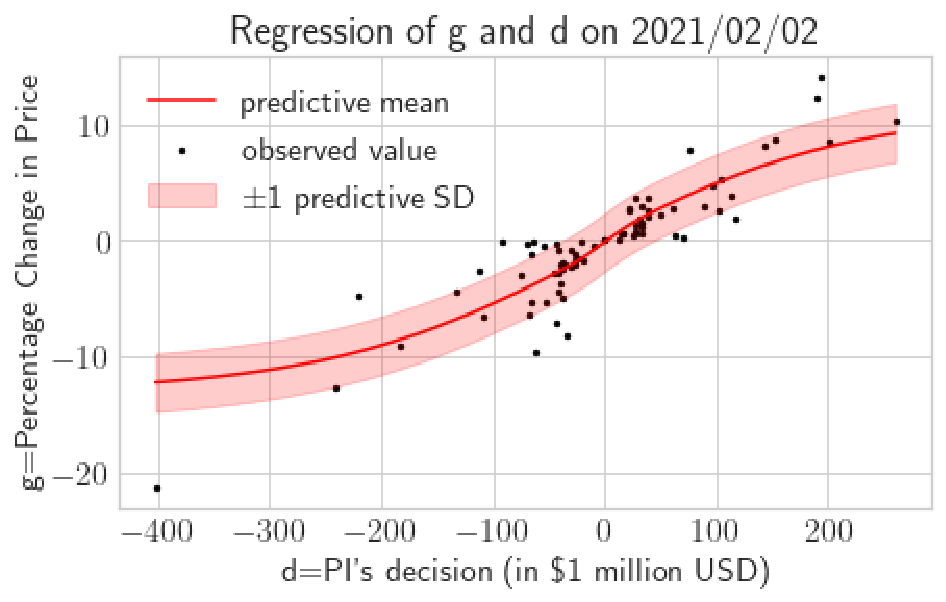}

}

\caption{Regression with $\sigma=2.5$ and $\lambda=0.7$\label{fig:RegressionSet1}}
\end{figure}

\begin{figure}
\subfloat[\label{fig:3a}]{\includegraphics[width=4.5cm,height=2.5cm]{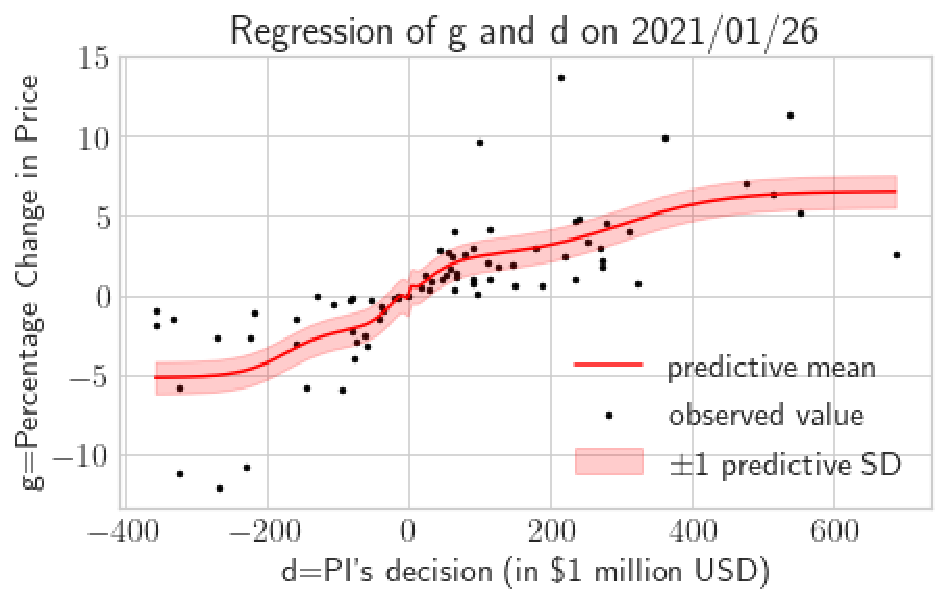}}\subfloat[\label{fig:3b}]{\includegraphics[width=4.5cm,height=2.5cm]{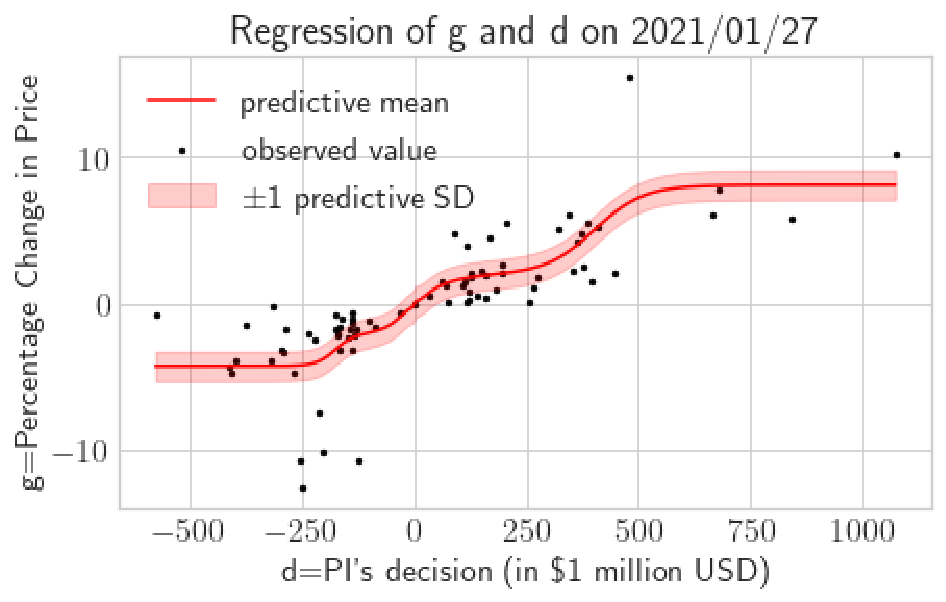}}\medskip{}

\subfloat[\label{fig:3c}]{\includegraphics[width=4.5cm,height=2.5cm]{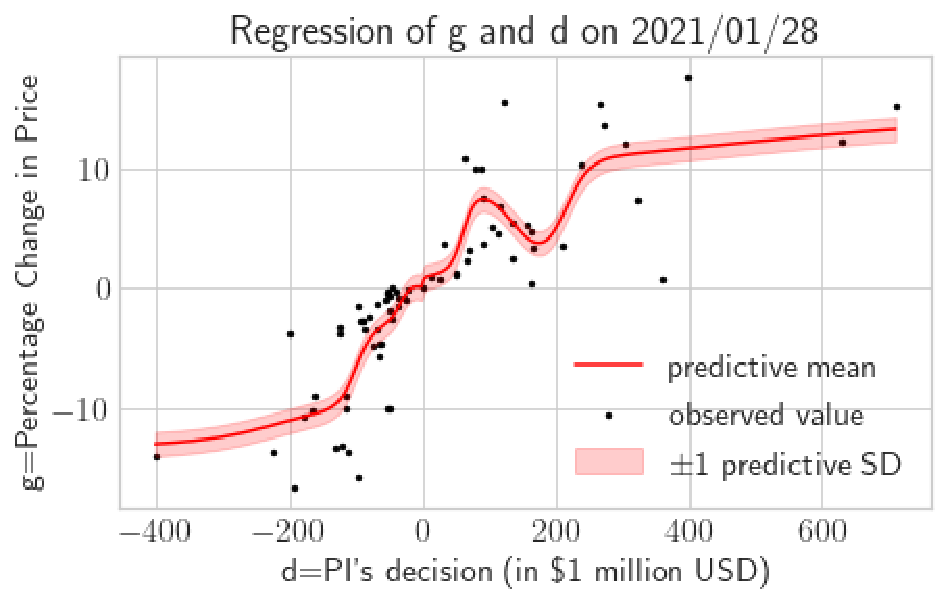}}\subfloat[\label{fig:3d}]{\includegraphics[width=4.5cm,height=2.5cm]{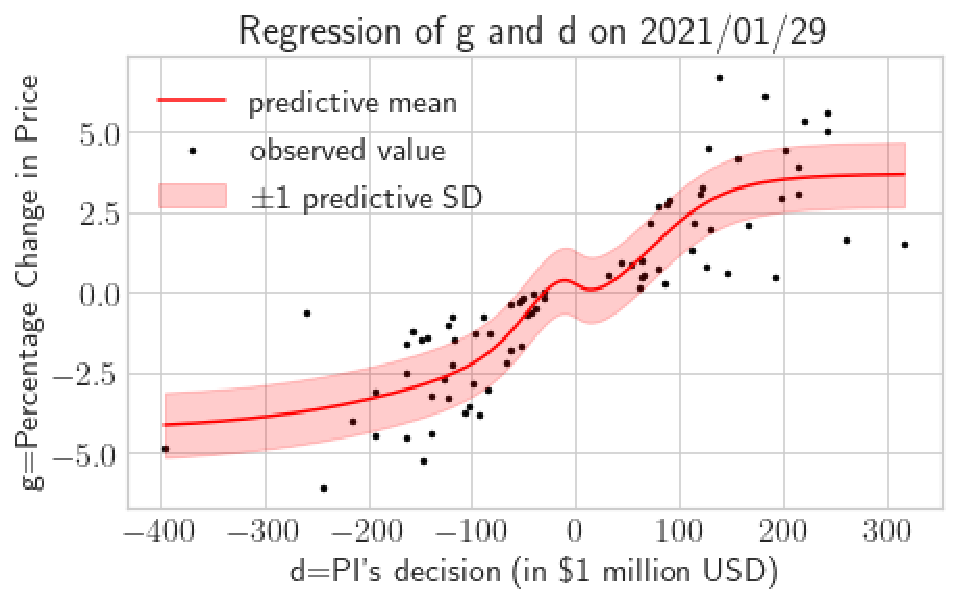}}\medskip{}

\subfloat[\label{fig:3e}]{\includegraphics[width=4.5cm,height=2.5cm]{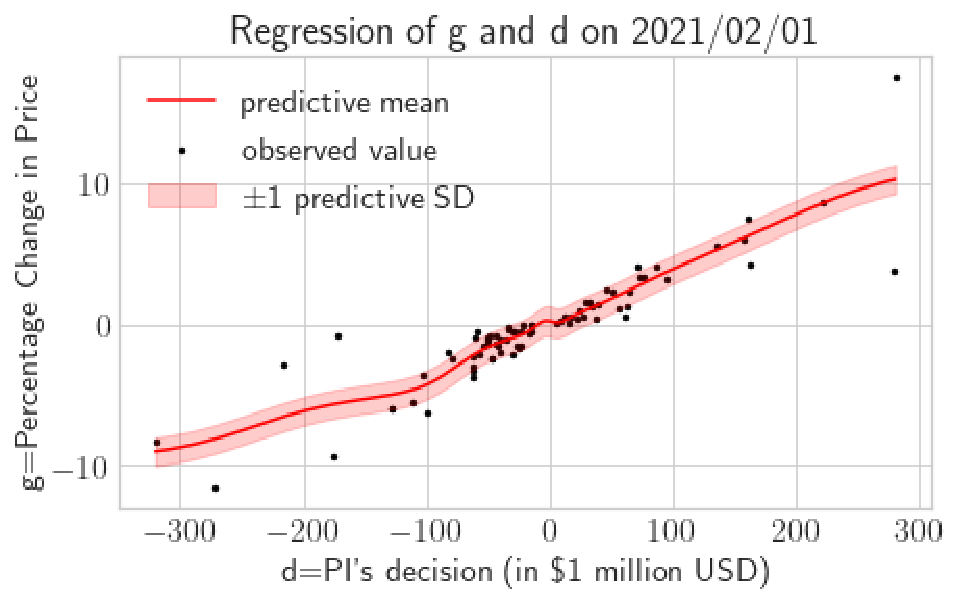}}\subfloat[\label{fig:3f}]{\includegraphics[width=4.5cm,height=2.5cm]{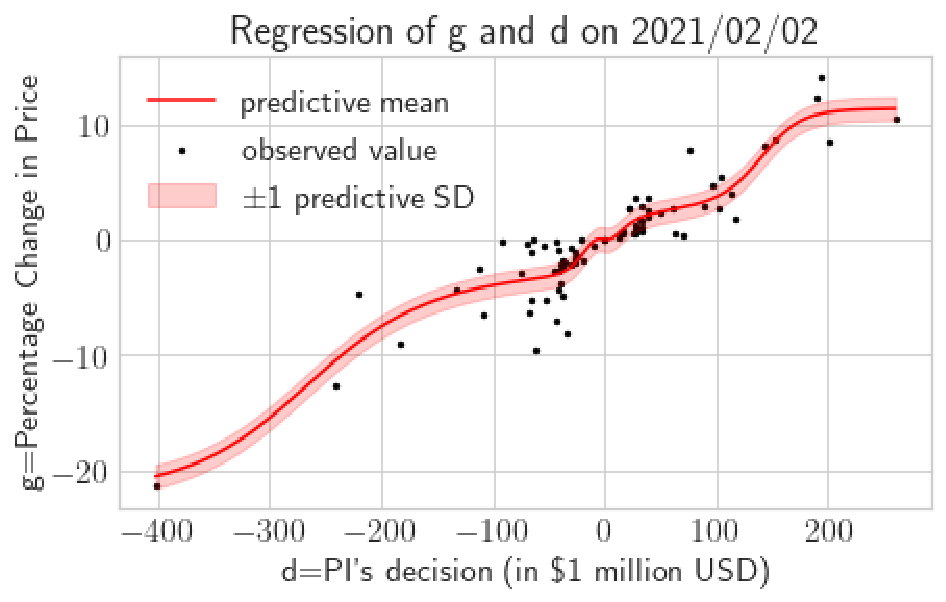}

}

\caption{Regression with $\sigma=1$, $\lambda=0.1$\label{fig:RegressionSet2}}
\end{figure}

We first train B-DNN respectively on an episode basis with the training
dataset $\mathcal{\mathscr{D}}_{i,i=1\ldots6}$ to determine the posterior
distribution over the weights and then work out the posterior predictive
distribution over price change led by testing samples of quantized
PI's decision. The testing samples are produced to be evenly spaced
over the interval between the largest and smallest training sample
of quantized PI's decision in each episode. There are two fine-tuned
hyperparameters $\sigma$ and $\lambda$ for optimizing our model's
performance. Since $\sigma$ has a significant influence on posterior
predictive variance as shown in Equation (\ref{eq:predictive distri of g*}),
the optimal value of $\sigma$ can be found out by the \textquotedbl 68--95--99.7
rule\textquotedbl . We carry out Bayesian inference and regression
analysis with two sets of parameters ($\sigma=2.5$, $\lambda=0.7$,
and $\sigma=1$, $\lambda=0.1$) and exhibits the results in turn
by Figure \ref{fig:RegressionSet1} and Figure \ref{fig:RegressionSet2}.
It is observed that $\sigma=2.5$ offers a fitting posterior predictive
variance to satisfy in principle the \textquotedbl 68--95--99.7
rule\textquotedbl , while $\sigma=1$ leads to a too small posterior
predictive variance to match the actual data distribution. The experimental
evaluation also demonstrates that the overfitting is effectively prevented
by $\lambda$. As shown in Figure \ref{fig:RegressionSet1}, the overfitting
is fully suppressed by a relatively large regularization parameter
$\lambda=0.7$. On the other hand, overfitting emerges when the regularization
parameter decreases to a smaller value, for instance, $\lambda=0.1$
as illustrated in Figure \ref{fig:RegressionSet2}. Furthermore, we
can observe from the numeric results that the stock price change has
an approximately positive correlation with quantized PI's decision.

\section{Inverse PI's decision from visible and hidden states\label{sec:Inverse-PI's-decision}}

The previous section presents a DNN-based approach to determine the
state transition (i.e., price change) after executing an action (i.e.,
principal investor's decision). We still wonder how an action is made
and especially how to associate an action with the current state (i.e.,
open price). Answering this query equals the proffer of the last remaining
piece of our MDP. This section will find out the underlying law to
simulate PI's successive decision making. 

When we ground the numerical results on the real equities pricing
data in the previous section, it is observed that higher quantized
PI's decision approximately leads to higher stock price change. This
phenomenon can be theoretically explained by the Keynesian beauty
contest \cite{Keynes1,Keynes2}. Keynesian beauty contest describes
a beauty contest where voters win a prize for voting the most popular
faces among all voters. Under such a rule of the contest, a voter
tends to guess the other voters' choices and match the choice of the
majority based on his or her own guess. Keynes believed that the similar
behavior pattern applies to the investors within the stock market.
Most investors attempt to make the same decision that the majority
do rather than build a decision on their own evaluation of fundamental
profitability. However, it is very hard to guess another investor's
concurrent decision for a retail investor. For this reason, the stock
market evolves into a leader-follower type Keynesian beauty contest
where a single retail investor (i.e., follower) tends to follow the
instantaneous observation of PI's decision and thinks it as the most
investors' concurrent decision. The leader-follower type Keynesian
beauty contest explains well the approximately positive correlation
between PI's decision and the stock price change. Nevertheless, PI's
decision making is still inadequate to be inferred from the leader-follower
type Keynesian beauty contest.

We believe that two types of factors influence PI's decision making:
visible and hidden. The visible factor is observable, while the hidden
factor is hard to be quantized or detected, for example, the impact
of news on investors\textquoteright{} anticipation of the investment
return and thus on their investment behavior. 

\begin{figure}
\includegraphics[scale=0.7]{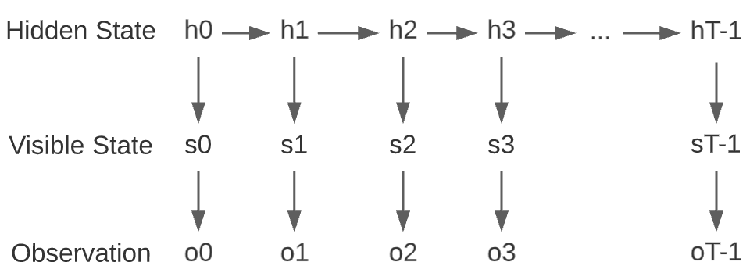}

\caption{Visible-hidden Markov network\label{fig:Hidden-visible-Markov-network}}
\end{figure}

The conventional MDP applies a policy function to map state space
to action space. However, the policy function is not sufficient for
mapping two state spaces to action space. In order to identify the
dependency of PI's decision making on visible factor and hidden factor,
we build a visible-hidden Markov network as shown in Figure \ref{fig:Hidden-visible-Markov-network}.
The hidden factor is modeled by a random variable (i.e., hidden state)
which changes through time and the visible factor is modeled by another
random variable (i.e., visible state). The hidden state transits over
time slots, and its transition is a Markov process. In the same time
slot, the hidden state, the visible state, and the random variable
representing PI's decision (i.e., observation) are pairwise correlated.
We design a chain for connecting these random variables to enable
the calculation of the dependencies among them. This chain design
differentiates the visible-hidden Markov network from hidden Markov
models \cite{HiddenMarkovModel} which considers only two coinstantaneous
components without the visible state. In each time slot, the hidden
state is directly connected to the visible state and then indirectly
chained to the observation via the visible state. 

We specify the notations used in the visible-hidden Markov network
below: 

$\zeta=\left\{ \zeta_{0},\zeta_{1},\ldots,\zeta_{J-1}\right\} $:
hidden state sample space,

$\nu=\left\{ \nu{}_{0},\nu_{1},\ldots,\nu_{K-1}\right\} $: visible
state sample space,

$\omega=\left\{ \omega_{0},\omega_{1},\ldots,\omega_{L-1}\right\} $:
observation sample space,

$Z=\left\{ z_{0},z_{1},\ldots,z_{J-1}\right\} $: initial hidden state
distribution,

$\mathbf{A}=\left\{ a_{ij}\right\} \in\mathbb{R}^{J\times J}$: hidden
state transition matrix, 

$a_{ij}=P\left\{ \mathrm{hidden\;state}\;\zeta_{j}\;\mathrm{at}\;t+1\mid\mathrm{hidden\;state}\;\zeta_{i}\;\mathrm{at}\;t\right\} $,

$\mathbf{B}=\left\{ b_{i}\left(k\right)\right\} \in\mathbb{R}^{J\times K}$:
visible state probability matrix,

$b_{i}\left(k\right)=P\left\{ \mathrm{visible\;state}\;v_{k}\;\mathrm{at}\;t\mid\mathrm{hidden\;state}\;\zeta_{i}\;\mathrm{at}\;t\right\} $,

$\mathbf{C}=\left\{ c_{k}\left(l\right)\right\} \in\mathbb{R}^{K\times L}$:
observation probability matrix,

$c_{k}\left(l\right)=P\left\{ \mathrm{observation\;}o_{l}\;\mathrm{at}\;t\mid\mathrm{visible\;state}\;v_{k}\;\mathrm{at}\;t\right\} $, 

$H=\left\{ h_{0},h_{1},\ldots,h_{\mathrm{T}-1}\right\} $: hidden
state sequence,

$S=\left\{ s_{0},s_{1},\ldots,s_{\mathrm{T}-1}\right\} $: visible
state sequence,

$O=\left\{ o_{0},o_{1},\ldots,o_{\mathrm{T}-1}\right\} $: observation
sequence,

$\Gamma=\left\{ \mathbf{A},\mathbf{B},\mathbf{C},Z\right\} $: Visible-hidden
Markov network.

In our scenario, the hidden state indicates bullish or bearish news
received just before a time slot, so the hidden state sample space
$\zeta$ is with a size of 2. The visible state and the observation
chained to the same hidden state respectively represent the opening
price and quantized PI's decision in the same time slot influenced
by the just acquired market information. The visible state sample
space $\nu$ and the observation sample space $\omega$ are finite
and discrete compared with the continuous sample space of the opening
price and quantized PI's decision. For this reason, we borrow the
concept of the quantization in digital signal processing to map values
from the sample space of the opening price and quantized PI's decision
to $\nu$ and $\omega$:

\begin{equation}
\bar{s}=\lfloor\frac{s}{\frac{s_{max}-s_{min}}{\mathsf{Q}}}\rfloor,\label{eq:quatization}
\end{equation}
where $\left\lfloor \right\rfloor $ is the floor function. $s$ stands
for a sample value of the opening price or quantized PI's decision.
$s_{max}$ and $s_{min}$ mark respectively the upper and lower limits
of the sample space of the opening price or quantized PI's decision.
$\mathsf{Q}$ denotes the size of $\nu$ or $\omega$ which is $K$
or $L$. $\bar{s}$ is mapped from $s$ to constitute $\nu$ or $\omega$.

The hidden state is initialized from a distribution $Z$ and transits
based on a transition probability matrix $\mathbf{A}$. In the same
time slot, the visible state is related to the hidden state by a probability
matrix $\mathbf{B}$ and the observation is associated with the visible
state by another probability matrix $\mathbf{C}$ to form a chain.
Our objective is to find $\Gamma=\left\{ \mathbf{A},\mathbf{B},\mathbf{C},Z\right\} $
which best possible fits the visible state sequence $S$ and the observation
sequence $O$. To this end, an algorithm is proposed to modulate $\Gamma$
by iteratively increasing $P\left(S,O\mid\Gamma\right)$:

Step 1: Initialize $\Gamma$ and set $P\left(S,O\mid\Gamma\right)=0$.

Step 2: Calculate the conditional joint probability function $\alpha_{t}\left(i\right)$,
$\beta_{t}\left(i\right)$,$\gamma_{t}\left(i,j\right)$ and the conditional
probability function $\gamma_{t}\left(i\right)$: 

\begin{align}
\alpha_{t=0}\left(i\right) & =P\left(\mathcal{\mathit{s}}_{0},\mathcal{\mathit{o}}_{0},h_{0}=\zeta_{i}|\Gamma\right)\nonumber \\
 & =z_{i}b_{i}\left(s_{0}\right)c_{s_{0}}\left(o_{0}\right),\label{eq:alpha0}
\end{align}

\begin{align}
\alpha_{t>0}\left(i\right) & =P\left(\mathcal{\mathit{s}}_{0},\mathcal{\mathit{s}}_{1},\ldots,\mathcal{\mathit{s}}_{t},\mathcal{\mathit{o}}_{0},\mathcal{\mathit{o}}_{1},\ldots,o_{t},h_{t}=\zeta_{i}|\Gamma\right)\nonumber \\
 & =\sum_{j=0}^{K-1}\alpha_{t-1}\left(j\right)a_{ji}b_{i}\left(s_{t}\right)c_{s_{t}}\left(o_{t}\right),\label{eq:alpha}
\end{align}

\begin{equation}
\beta_{t=\mathrm{T}-1}\left(i\right)=P\left(\mathcal{\mathit{s}}_{\mathrm{T}-1},\mathcal{\mathit{o}}_{\mathrm{T}-1}|h_{\mathrm{T}-1}=\zeta_{i},\Gamma\right)=1,\label{eq:betaT-1}
\end{equation}

\begin{align}
\beta_{t<\mathrm{T}-1}\left(i\right)\nonumber \\
=P\left(\mathcal{\mathit{s}}_{t+1},\mathcal{\mathit{s}}_{t+2},\ldots,\mathcal{\mathit{s}}_{T-1},\mathcal{\mathit{o}}_{t+1},\mathcal{\mathit{o}}_{t+2},\ldots,\mathcal{\mathit{o}}_{T-1}|h_{t}=\zeta_{i},\Gamma\right)\nonumber \\
=\sum_{j=0}^{K-1}a_{ij}b_{j}\left(s_{t+1}\right)c_{s_{t+1}}\left(o_{t+1}\right)\beta_{t+1}\left(j\right) & ,\label{eq:beta t>0}
\end{align}

\begin{align}
\gamma_{t=0\ldots\mathrm{T}-2}\left(i,j\right) & =P\left(h_{t}=\zeta_{i},h_{t+1}=\zeta_{j}\mid S,O,\Gamma\right)\nonumber \\
 & =\frac{\alpha_{t}\left(i\right)a_{ij}b_{j}\left(s_{t+1}\right)c_{s_{t+1}}\left(o_{t+1}\right)\beta_{t+1}\left(j\right)}{P\left(S,O\mid\Gamma\right)}\nonumber \\
 & =\frac{\alpha_{t}\left(i\right)a_{ij}b_{j}\left(s_{t+1}\right)c_{s_{t+1}}\left(o_{t+1}\right)\beta_{t+1}\left(j\right)}{\sum_{i=0}^{J-1}\alpha_{t=\mathrm{T}-1}\left(i\right)},\label{eq:gammaij}
\end{align}

\begin{align}
\gamma_{t=\mathrm{T}-1}\left(i\right) & =P\left(h_{t}=\zeta_{i}\mid S,O,\Gamma\right)\nonumber \\
 & =\frac{\alpha_{t=0}\left(i\right)}{\sum_{i=0}^{J-1}\alpha_{t=0}\left(i\right)},\label{eq:gmmaiT-1}
\end{align}

\begin{align}
\gamma_{t=0\ldots\mathrm{T}-2} & \left(i\right)=P\left(h_{t}=\zeta_{i}\mid S,O,\Gamma\right)\nonumber \\
 & =\sum_{j=0}^{J-1}\gamma_{t}\left(i,j\right).\label{eq:gammai}
\end{align}

\begin{figure}
\subfloat[\label{fig:5a}]{\includegraphics[width=4.5cm,height=3.5cm]{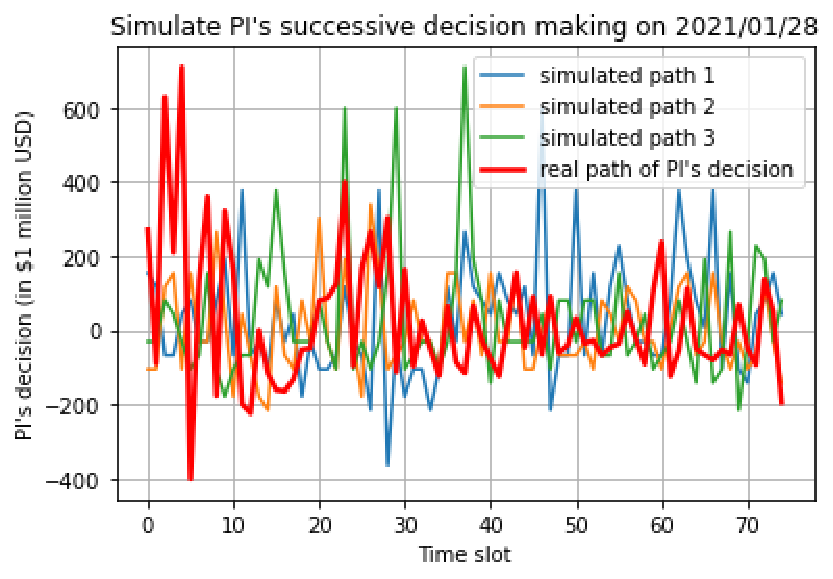}}\subfloat[\label{fig:5b}]{\includegraphics[width=4.5cm,height=3.5cm]{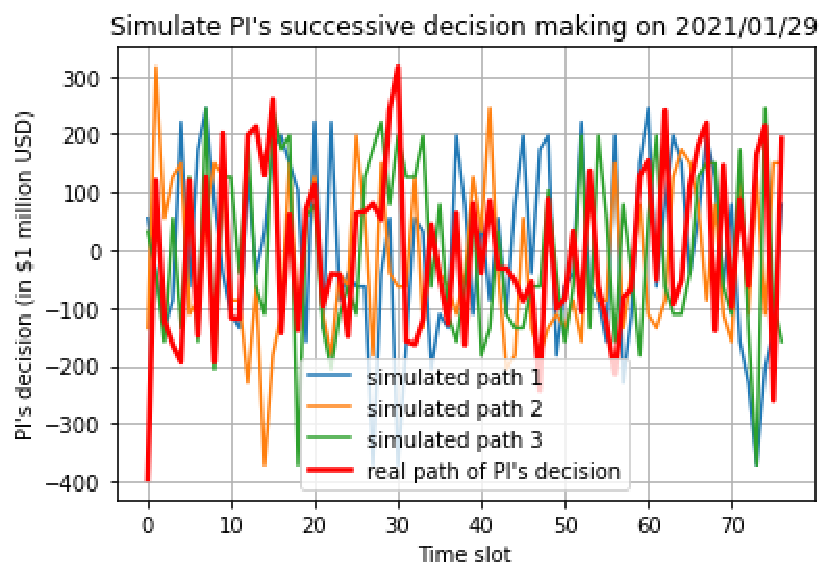}}\caption{Simulation of PI's successive decision making\label{fig:PIDecision}}
\end{figure}

\begin{figure}
\subfloat[\label{fig:6a}]{\includegraphics[width=4.5cm,height=3.5cm]{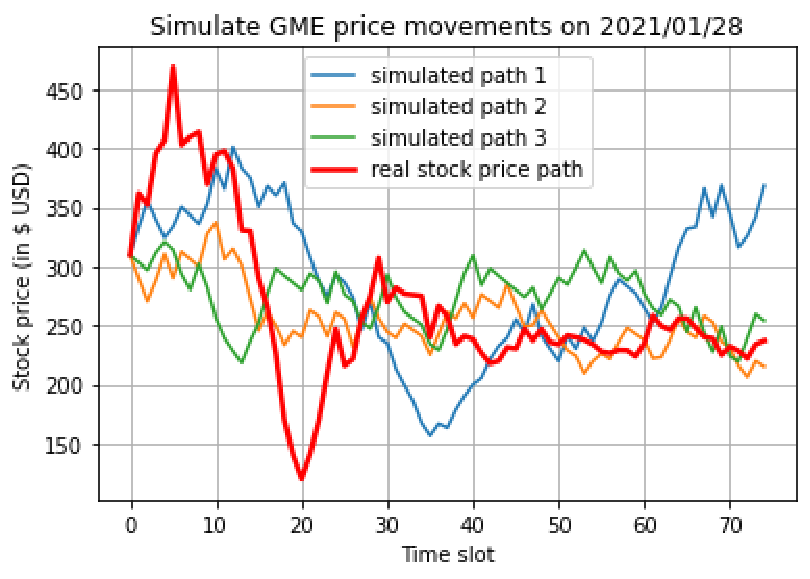}}\subfloat[\label{fig:6b}]{\includegraphics[width=4.5cm,height=3.5cm]{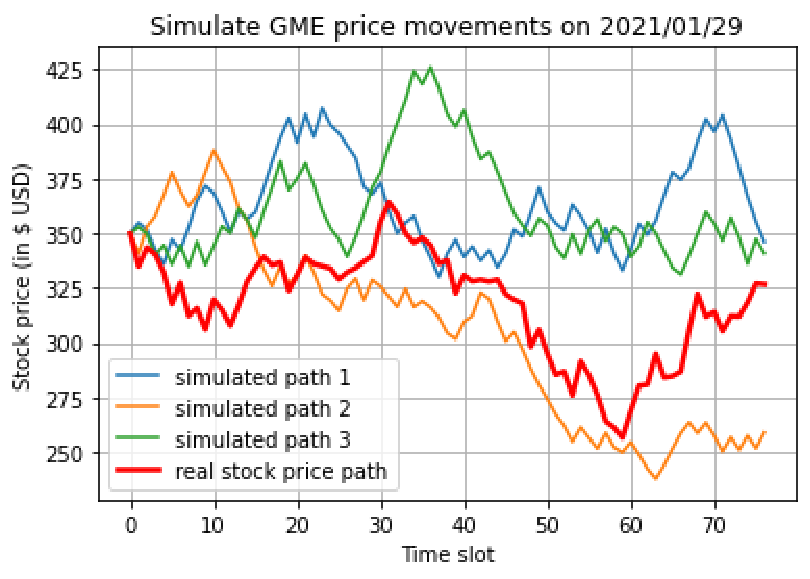}}\caption{Simulation of GME price movements\label{fig:priceMovements}}
\end{figure}

Step 3: Re-estimate $\Gamma$:

\begin{equation}
z_{i}=\gamma_{t=0}\left(i\right),\label{eq:zi}
\end{equation}

\begin{equation}
a_{ij}=\frac{\sum_{t=0}^{T-2}\gamma_{t}\left(i,j\right)}{\sum_{t=0}^{T-2}\gamma_{t}\left(i\right)},\label{eq:aij}
\end{equation}

\begin{equation}
b_{i}\left(k\right)=\frac{\sum_{t\in\left\{ 0,1,\ldots,T-1\right\} ,s_{t}=\nu{}_{k}}\gamma_{t}\left(i\right)}{\sum_{t=0}^{T-1}\gamma_{t}\left(i\right)},\label{eq:bik}
\end{equation}

\begin{equation}
c_{k}\left(l\right)=\frac{\sum_{t\in\left\{ 0,1,\ldots,T-1\right\} ,s_{t}=\nu{}_{k},o_{t}=\omega_{l}}\gamma_{t}\left(i\right)}{\sum_{t=0}^{T-1}\gamma_{t}\left(i\right)}.\label{eq:ckl}
\end{equation}

Step 4: If $P\left(S,O\mid\Gamma\right)=\sum_{i=0}^{J-1}\alpha_{t=\mathrm{T}-1}\left(i\right)$
climbs, go to Step 2, otherwise end the loop.

The parameters of a visible-hidden Markov network are initialized
before entering the iterative process and then updated in each iteration
until $P\left(S,O\mid\Gamma\right)$ doesn't grow anymore. $\mathbf{A}$,
$\mathbf{B}$, $\mathbf{C}$, and $Z$ are row stochastic, moreover,
their rows respectively belong to a standard $J-1$, $J-1$, $K-1$,
and $L-1$ simplex which are the support of the Dirichlet distribution
of order $J-1$, $J-1$, $K-1$, and $L-1$. Therefore, we can employ
the Dirichlet distribution to randomize $\mathbf{A}$, $\mathbf{B}$,
$\mathbf{C}$, and $Z$ to keep away from a local maximum which prevents
the evolution of the model.

Figure \ref{fig:1c} and \ref{fig:1d} clearly show that PI's decision
making strategy varies from one episode to another: PI preferred to
give a decision with large absolute value at the early stage on Jan.
28, 2021, on the other hand, tended to spread large trades over time
on Jan. 29, 2021. In light of these signs, our determination of a
visible-hidden Markov network is episode-based. We measure $\Gamma$
in each episode to generate a time series of the probable evolution
of PI's decision making and further transform them to stock price
paths through Bayesian inference and regression analysis.

We imitate PI's successive decision making based on observations from
the training dataset $\mathcal{\mathscr{D}}_{i,i=1\ldots6}$. The
hyperparameters are given the following values: $J=2$, and $K=L=30$.
We parametrize Dirichlet distribution with $\alpha_{i}=1000$, so
that the initial $\mathbf{A}$, $\mathbf{B}$, $\mathbf{C}$, and
$Z$ are row stochastic and have near uniform values, where $\alpha_{i}$
denotes any component of the parameter vector $\boldsymbol{\alpha}$.
Figure \ref{fig:PIDecision} exhibits three simulated paths of PI's
successive decision making along with the real evolution of PI's decision.
In view of article space, we only show the experimental results in
two representative episodes for comparison in figures. According to
Figure \ref{fig:5a} and \ref{fig:5b}, we find that the simulated
paths on Jan. 29, 2021 are more volatile than those on Jan. 28, 2021.
This correctly reflects the discrepancy of the volatility of real
PI's decision on these two days and thus well proves the effectiveness
of our algorithm. The B-DNN is trained by the training dataset $\mathcal{\mathscr{D}}_{i,i=1\ldots6}$
when we perform Bayesian inference in the section \ref{sec:Effect-of-the}.
We continue to carry out regression analysis grounded on this trained
B-DNN for each evolutionary path of PI's decision. For the sake of
simplicity, the price change is simulated as its posterior predictive
mean. Figure \ref{fig:priceMovements} illustrates our simulated price
movements of the underlying stock together with the real price path.

\section{Reinforcement learn option price ground upon imitative PI's chronological
decision making}

Our algorithm introduced in the section \ref{sec:Inverse-PI's-decision}
probabilistically imitates PI's chronological decision making. With
the help of B-DNN inference-regression presented in the section \ref{sec:Effect-of-the},
likely evolutionary paths of PI's decision are transformed into probable
price paths of the underlying stock. These simulated price paths of
the underlying stock yield the cross-sectional information for optimizing
a dynamically hedged portfolio. This section expounds how to reinforcement
learn option price using the cross-sectional information stemmed from
imitative PI's chronological decision making. 

Suppose that we sell a European option with the terminal payoff $\mathsf{H}\left(S_{\mathsf{T}}\right)$,
where $S_{\mathsf{T}}$ denotes the underlying stock price at expiry
date $\mathsf{T}$. For the sake of the suitability of our algorithm
for both calls and puts, switching the below expression of the terminal
payoff will turn pricing/hedging a call option into a put option:
\begin{equation}
\mathsf{H}\left(S_{\mathsf{T}}\right)=\begin{cases}
\max\left(S_{\mathsf{T}}-K,0\right),\mathrm{if\;calls} & \mathrm{}\\
\max\left(K-S_{\mathsf{T}},0\right),\mathrm{if\;puts}.
\end{cases}\label{eq:terminal payoff}
\end{equation}

With the proceeds from the sale of the option, we build a portfolio
$\Pi_{t}$ which consists of $a_{t}$ units of the stock with price
$S_{t}$ and a deposit in the risk-free bank account $B_{t}$. With
the aim of dynamically hedging the option, we rebalance this portfolio
to keep its value replicating the value of the option at time $t\leqslant\mathsf{T}$:

\begin{equation}
\Pi_{t}=a{}_{t}S_{t}+B_{t}.\label{eq:value of the option}
\end{equation}
At maturity we close out the position as we no longer need to hedge
the option. Consequently, the terminal payoff is the portfolio value
at maturity:

\begin{equation}
\Pi_{\mathsf{T}}=\mathsf{H}\left(S_{\mathsf{T}}\right)=B_{t}.\label{eq:terminal portfolio value}
\end{equation}

We ignore transaction costs or other frictions in this paper due to
the fact that low rates or fixed commissions are offered to the customers
with high volume of trades executed (e.g., High-Frequency Trading
(HFT) firms). We force our dynamically hedged portfolio $\Pi_{t}$
to be self-financing in such a manner that there is no external infusion
or withdrawal of cash over the lifetime of the option, and the spread
of portfolio values at different times comes from the continuously
compounded risk-free interest rate $r$ and the change in underlying
stock price $\bigtriangleup S_{t}$ :

\begin{equation}
\Pi_{t+1}-e^{r}\Pi_{t}=a_{t}\bigtriangleup S_{t}+\kappa f_{\boldsymbol{\theta}^{*}}\left(F\right)a_{t}S_{t},\label{eq: portfolio values difference}
\end{equation}
where $\bigtriangleup S_{t}=S_{t+1}-e^{r}S_{t}$ and $F=\left(a_{t+1}-a_{t}\right)\varTheta S_{t}$.

The cross-sectional information $\mathcal{F}\left(t\right)=\left\{ S_{t}^{1},\ldots,S_{t}^{U}\right\} $
collects possible prices at time $t$ from all the simulated price
paths, where $U$ is the number of simulated paths and $S_{t}^{u},_{u=1,\ldots,U}$
denotes the price at time $t$ sampled from the $u$th path. Our cross-sectional
information is derived from the existing PI's decision, so we design
a compensation term $\kappa f_{\boldsymbol{\theta}^{*}}\left(F\right)a_{t}S_{t}$
to put in the price change caused by our hedging activity if we are
also referred to as PI. $f_{\boldsymbol{\theta}^{*}}\left(F\right)$
maps a large-scale volume of buying or selling the underlying stock
to a sharp rise or drop in price via our trained B-DNN. We may sell
multiple options covering a significant amount of shares of the underlying
stock. We use $\varTheta$ to denote the total number of shares of
the underlying stock. The coefficient $\kappa$ gives a degree of
freedom to scale the compensation term. If we are a single retail
investor who sells a mini option, then this compensation term vanishes
as $\kappa$ tends to 0. Monte Carlo sample-based reinforcement learning
\cite{QLBS} uses the cross-sectional information which is generated
by sampling a pre-set probability distribution. Compared with Monte
Carlo sample-based reinforcement learning, our algorithm can better
fit into each circumstance as we adopt the cross-sectional information
that reflects the existing PI's decision and add an compensation term
to compensate for the effect of our hedging activity on the price
change.

When price an option at any time within its lifetime, the risk of
running out cash to continue rebalancing the portfolio due to the
fluctuation in the market price of the underlying stock should be
taken into account. To this end, the ask price $\mathcal{C}_{t}$
at time $t\leqslant\mathsf{T}$ is designed to be the sum of the risk
free price and the risk-adjustment weighted by the risk-aversion coefficient
$\eta$ \cite{QLBS} as expressed in Equation (\ref{eq:option price}).
The risk free price is the expected value of the portfolio $\Pi_{t}$.
The risk-adjustment accumulates the risk from time $t$ until the
expiry date $\mathsf{T}$. The risk at time $t$ is quantized as the
expected discounted variance of the portfolio $\Pi_{t}$.

\begin{equation}
\mathcal{C}_{t}\left(S_{t}\right)=\mathbb{E}_{t}\left[\Pi_{t}+\eta\sum_{t^{\prime}=t}^{\mathsf{T}}e^{-r\left(t^{\prime}-t\right)}\mathrm{Var}\left[\Pi_{t^{\prime}}\mid\mathcal{F}_{t^{\prime}}\right]|\mathcal{F}_{t}\right].\label{eq:option price}
\end{equation}

Geometric Brownian motion is the most commonly used model of stock
price behavior. By applying Itô's lemma to the random variable of
stock price which follows Geometric Brownian motion, the stock price
at time $t$ has been proved to be lognormally distributed \cite{J. Hull}: 

\begin{equation}
\ln S_{t}\sim\mathcal{N}\left(\ln S_{0}+\left(\mu-\frac{\sigma_{s}^{2}}{2}\right)t,\sigma_{s}^{2}t\right),\label{eq:lognormal distributed}
\end{equation}
where $\mu$ is the expected rate of return per time-slot from the
stock, and $\sigma_{s}$ is the volatility of the stock price.

Equation (\ref{eq:lognormal distributed}) implies $S_{t}$ is not
martingale due to the constant drift rate $\mu-\frac{\sigma_{s}^{2}}{2}$.
When price the option, we prefer the calculation to be based on an
independent variable without constant drift. Therefore, we expand
the independent variable $S_{t}$ in Equation (\ref{eq:value of the option}),
(\ref{eq: portfolio values difference}), and (\ref{eq:option price})
to a function in terms of an independent variable without constant
drift $\widetilde{S}_{t}$:

\begin{equation}
S_{t}=\exp\left(\widetilde{S}_{t}+\left(\mu-\frac{\sigma_{s}^{2}}{2}\right)t\right).\label{eq:variable without drift}
\end{equation}
$S_{t}$ afterwards will signify the expansion of stock price in terms
of $\widetilde{S}_{t}$ in all equations.

The following notations will denote for short the other variables
which are dependent on $\widetilde{S}_{t}$ in subsequent derivations: 

\begin{equation}
\widetilde{a}_{t}=\widetilde{a}_{t}\left(\widetilde{S}_{t}\right)=a_{t},\label{eq:action}
\end{equation}

\begin{equation}
\pi=\pi\left(\widetilde{S}_{t},t\right)=\widetilde{a}_{t},\label{eq:policy}
\end{equation}

\begin{equation}
\bigtriangleup S_{t}=e^{\widetilde{S}_{t+1}+\left(\mu-\frac{\sigma_{s}^{2}}{2}\right)\left(t+1\right)}-e^{r}e^{\widetilde{S}_{t}+\left(\mu-\frac{\sigma_{s}^{2}}{2}\right)t},\label{eq:delta S}
\end{equation}

\begin{equation}
\Pi_{t}=\Pi_{t}\left(\widetilde{S}_{t}\right),\label{eq:portfolio value}
\end{equation}
where $\pi$ is the policy function which maps $\widetilde{S}_{t}$
and $t$ to the position in the stock at time $t$.

We are aiming to find a policy to price the option with the satisfaction
for the only necessary demand of hedging risk, or other kind of statement,
catch the optimal or nearly optimal $\pi$ that minimizes the ask
price $\mathcal{C}_{t}\left(\widetilde{S}_{t}\right)$. Our goal can
be achieved by using reinforcement learning which translates the aim
into learning $\pi$ that maximizes the expected cumulative reward
via maximizing the state-value function and the action-value function
\cite{Reinforcement Learning: An Introduction}. Reinforcement learning
is found on a MDP. In our scenario, the stock price $\widetilde{S}_{t}\in\mathcal{S}$
is the state and the position in the stock $\widetilde{a}_{t}\in\mathcal{A}$
is the action. Accordingly, the sample space of the stock price $\mathcal{S}$
and the position $\mathcal{A}$ respectively form the state space
and the action space. The state-value function for a policy $\pi$
at time $t$ is denoted as $V_{t}^{\pi}\left(\widetilde{S}_{t}\right)$
and it is designed as the negative of the option price. Equation (\ref{eq: portfolio values difference})
formulates the chronological recursive relationship that expresses
the portfolio's value at time $t$ in respect of its value at the
subsequent time $t+1$. We can exploit this chronological recursive
relationship to find the Bellman equation for the state-value function
and the reward: 

\begin{align}
V_{t}^{\pi}\left(\widetilde{S}_{t}\right)=-\mathcal{C}_{t}\left(\widetilde{S}_{t}\right)\label{eq:-C}\\
=\mathbb{E}_{t}\left[-\Pi_{t}-\eta\sum_{t^{\prime}=t}^{\mathsf{T}}e^{-r\left(t^{\prime}-t\right)}\mathrm{Var}\left[\Pi_{t^{\prime}}\mid\mathcal{F}_{t^{\prime}}\right]|\mathcal{F}_{t}\right]\label{eq:-C2}\\
=\mathbb{E}_{t}\left[-\Pi_{t}-\eta\mathrm{Var}\left[\Pi_{t^{\prime}}\right]-\eta\sum_{t^{\prime}=t+1}^{\mathsf{T}}e^{-r\left(t^{\prime}-t\right)}\mathrm{Var}\left[\Pi_{t^{\prime}}\mid\mathcal{F}_{t^{\prime}}\right]|\mathcal{F}_{t}\right]\label{eq:-C3}\\
=\mathbb{E}_{t}\left[R_{t}\left(\widetilde{S}_{t},\widetilde{a}_{t},\widetilde{S}_{t+1}\right)+\gamma V_{t+1}^{\pi}\left(\widetilde{S}_{t+1}\right)\right] & .\label{eq:Bellman}
\end{align}
Equation (\ref{eq:-C2}) is obtained by plugging Equation (\ref{eq:option price})
into Equation (\ref{eq:-C}). Equation (\ref{eq:Bellman}) is the
Bellman equation for $V_{t}^{\pi}\left(\widetilde{S}_{t}\right)$,
where the reward $R_{t}\left(\widetilde{S}_{t},\widetilde{a}_{t},\widetilde{S}_{t+1}\right)$
is derived as below by using the chronological recursive relationship
given in Equation (\ref{eq: portfolio values difference}):

\begin{align}
R_{t}\left(\widetilde{S}_{t},\widetilde{a}_{t},\widetilde{S}_{t+1}\right) & =\gamma\left(\widetilde{a}_{t}\bigtriangleup S_{t}+\kappa f_{\boldsymbol{\theta}^{*}}\left(F\right)S_{t}\right)-\eta\mathrm{Var}\left[\Pi_{t}\mid\mathcal{F}_{t}\right]\nonumber \\
 & =\gamma\left(\widetilde{a}_{t}\bigtriangleup S_{t}+\kappa f_{\boldsymbol{\theta}^{*}}\left(F\right)S_{t}\right)-\eta\gamma^{2}\mathbb{E}_{t}\nonumber \\
 & \left[\left(\dot{\Pi}_{t+1}-\left(\widetilde{a}_{t}\bigtriangleup\dot{S}_{t}+\kappa f_{\boldsymbol{\theta}^{*}}\left(F\right)\widetilde{a}_{t}\dot{S}_{t}\right)\right)^{2}\right],\label{eq:Reward}
\end{align}
where $\gamma=e^{-r}$ is the discount rate, $\dot{\Pi}_{t+1}=\Pi_{t+1}-\overline{\Pi}{}_{t+1}$,
$\bigtriangleup\dot{S}_{t}=\bigtriangleup S_{t}-\bigtriangleup\overline{S}_{t}$,
$\dot{S}_{t}=S_{t}-\overline{S}_{t}$, and $\overline{\left(\cdot\right)}$
denotes the sample mean. Maximizing the reward is identical to maximizing
the cumulative risk-adjusted return as the last term of the reward
stands for the risk at time $t$ and the first term gives the quantized
portfolio's return between time $t$ and $t+1$. When $\eta$ tends
to infinity, the hedge becomes a pure risk hedge.

The action-value function ($Q$-function) returns the value of taking
action $a$ in state $s$ under a policy $\pi$ at time $t$. Consequently,
$Q$-function can be interpreted as a state-value function under the
condition of $\widetilde{S}_{t}=s$ and $\widetilde{a}_{t}=a$:

\begin{align}
Q_{t}^{\pi}\left(s,a\right) & =\mathbb{E}_{t}\left[-\Pi_{t}\mid\widetilde{S}_{t}=s,\widetilde{a}_{t}=a\right]\nonumber \\
 & -\mathbb{E}_{t}\left[\eta\sum_{t^{\prime}=t}^{\mathsf{T}}e^{-r\left(t^{\prime}-t\right)}\mathrm{Var}\left[\Pi_{t^{\prime}}\mid\mathcal{F}_{t^{\prime}}\right]|\widetilde{S}_{t}=s,\widetilde{a}_{t}=a\right].\label{eq:Q function}
\end{align}

We use $Q_{t}^{\pi^{\mathit{*}}}\left(\widetilde{S}_{t},\widetilde{a}_{t}^{*}\right)$
to stand for the $Q$-function value which is maximized by the optimal
policy $\pi^{\mathit{*}}$ which chooses the optimal action $\widetilde{a}_{t}^{*}$
in state $\widetilde{S}_{t}$ at time $t$:

\begin{align}
Q_{t}^{\pi^{\mathit{*}}}\left(\widetilde{S}_{t},\widetilde{a}_{t}^{*}\right) & =\underset{\pi}{\max}Q_{t}^{\pi}\left(\widetilde{S}_{t},\widetilde{a}_{t}\right)\nonumber \\
 & =\underset{\widetilde{a}_{t}\in\mathcal{A}}{\max}Q_{t}^{\pi}\left(\widetilde{S}_{t},\widetilde{a}_{t}\right),\label{eq:optimal Q}
\end{align}
where $\pi^{\mathit{*}}=\pi^{*}\left(\widetilde{S}_{t},t\right)$
and $\widetilde{a}_{t}^{*}=\widetilde{a}_{t}^{*}\left(\widetilde{S}_{t}\right)$
are both dependent on $\widetilde{S}_{t}$.

The Bellman optimality equation for the action-value function has
a similar structure to the Bellman equation for the state-value function
(Equation (\ref{eq:Bellman})) and can be expanded by plugging Equation
(\ref{eq:Reward}) into it:

\begin{align}
 & Q_{t}^{\pi}\left(\widetilde{S}_{t},\widetilde{a}_{t}\right)\nonumber \\
= & \mathbb{E}_{t}\left[R_{t}\left(\widetilde{S}_{t},\widetilde{a}_{t},\widetilde{S}_{t+1}\right)+\gamma Q_{t+1}^{\pi^{\mathit{*}}}\left(\widetilde{S}_{t+1},\widetilde{a}_{t+1}^{*}\right)\right]\label{eq:Q bellman optimality}\\
= & \gamma\mathbb{E}_{t}\left[\widetilde{a}_{t}\bigtriangleup S_{t}+\kappa f_{\boldsymbol{\theta}^{*}}\left(F\right)\widetilde{a}_{t}^{*}S_{t}+Q_{t+1}^{\pi^{\mathit{*}}}\left(\widetilde{S}_{t+1},\widetilde{a}_{t+1}^{*}\right)\right]\nonumber \\
 & -\eta\gamma^{2}\mathbb{E}_{t}\left[\left(\dot{\Pi}_{t+1}-\left(\widetilde{a}_{t}\bigtriangleup\dot{S}_{t}+\kappa f_{\boldsymbol{\theta}^{*}}\left(F\right)\widetilde{a}_{t}\dot{S}_{t}\right)\right)^{2}\right].\label{eq:Bellman optimality equation}
\end{align}

The terminal condition gives the optimal $Q$-function value at expiry
in terms of the terminal payoff:
\begin{equation}
Q_{\mathsf{T}}^{\pi^{\mathit{*}}}\left(\widetilde{S}_{\mathsf{T}}^{u},\widetilde{a}_{\mathsf{T}}^{*}\right)=-\mathsf{H}\left(S_{\mathsf{T}}\right)-\mathrm{Var}\left[\mathsf{H}\left(S_{\mathsf{T}}\right)\right],\label{eq:terminal condition}
\end{equation}
where $\widetilde{a}_{\mathsf{T}}^{*}=0$. 

Our purpose is to find $\pi^{*}\left(\widetilde{S}_{t},t\right)$
or $\widetilde{a}_{t}^{*}\left(\widetilde{S}_{t}\right)$, and $Q_{t}^{\pi^{\mathit{*}}}\left(\widetilde{S}_{t},\widetilde{a}_{t}^{*}\right)$.
However, the unknown value of $\widetilde{S}_{t}$ and the incalculable
expectation in the Bellman optimality equation prevent us from achieving
our goal. For the removal of the barrier, we use the cross-sectional
information $\widetilde{\mathcal{F}}=\left\{ \widetilde{S}_{t}^{u}\right\} _{u=1,\ldots,U}$
to take as many likely values of $\widetilde{S}_{t}$ into account
as possible, where $\widetilde{S}_{t}^{u}=\ln S_{t}^{u}-\left(\mu-\frac{\sigma_{s}^{2}}{2}\right)t$.
We believe that our simulated price paths have the same constant drift
as those following a lognormal distribution. Moreover, the incalculable
expectation in the Bellman optimality equation is approximated by
an empirical mean of $U$ observations.

In our scenario, the state space and the action space are both continuous.
We adopt basis functions, for example basis-splines, as an intermediary
for mapping two continuous spaces. A state sample $\widetilde{S}_{t}$
is first mapped to a set of basis functions and the optimal action
$\widetilde{a}_{t}^{*}$ is then expressed as a linear combination
of basis functions weighted by the optimal coefficients $w_{nt}^{*}$: 

\begin{equation}
\widetilde{a}_{t}^{*}=\sum_{n=1}^{\mathcal{B}}w_{nt}^{*}\psi_{n}\left(\widetilde{S}_{t}\right),\label{eq:basis1}
\end{equation}
where $\mathcal{B}$ represents the number of basis functions. Similarly,
the optimal $Q$-function value can be formulated as a linear combination
of basis functions weighted by the optimal coefficients $\varphi_{nt}^{*}$:

\begin{equation}
Q_{t}^{\pi^{\mathit{*}}}\left(\widetilde{S}_{t},\widetilde{a}_{t}^{*}\right)=\sum_{n=1}^{\mathcal{B}}\varphi_{nt}^{*}\psi_{n}\left(\widetilde{S}_{t}\right).\label{eq:basis2}
\end{equation}

Our objective of finding optimal policy and $Q$-function value is
consequently transformed into learning the optimal coefficients $w_{nt}^{*}$
and $\varphi_{nt}^{*}$. We elaborate the derivation of the following
closed-form solution of $w_{nt}^{*}$ in Appendix B:

\begin{equation}
w_{t}^{*}=\mathbf{E}_{t}^{-1}\mathbf{D}_{t},\label{eq:weight1}
\end{equation}

\begin{align}
E_{nm}^{t} & =\sum_{u=1}^{U}\left[\psi_{m}\left(\widetilde{S}_{t}^{u}\right)\bigtriangleup\dot{S}_{t}\varXi\sum_{n=1}^{\mathcal{B}}\psi_{n}\left(\widetilde{S}_{t}^{u}\right)\right]\nonumber \\
 & +\sum_{u=1}^{U}\left[\kappa f_{\boldsymbol{\theta}^{*}}\left(F\right)\psi_{m}\left(\widetilde{S}_{t}^{u}\right)\dot{S}_{t}\varXi\sum_{n=1}^{\mathcal{B}}\psi_{n}\left(\widetilde{S}_{t}^{u}\right)\right],\label{eq:EntryE}
\end{align}

\begin{equation}
D_{n}^{t}=\sum_{u=1}^{U}\left[\left(\frac{\bigtriangleup S_{t}^{u}}{2\eta\gamma}+\dot{\Pi}_{t+1}\varXi+\frac{\kappa f_{\boldsymbol{\theta}^{*}}\left(F\right)S_{t}^{u}}{2\eta\gamma}\right)\sum_{n=1}^{\mathcal{B}}\psi_{n}\left(\widetilde{S}_{t}^{u}\right)\right],\label{eq:entryD}
\end{equation}

\begin{equation}
\varXi=\left(\bigtriangleup\dot{S}_{t}+\kappa f_{\boldsymbol{\theta}^{*}}\left(F\right)\dot{S}_{t}\right),\label{eq:variable}
\end{equation}
where $w_{t}^{*}$ is a $\mathcal{B}$-dimensional vector with entry
$w_{nt}^{*}$, $\mathbf{E}_{t}$ is a matrix of size $\mathcal{B}\times\mathcal{B}$
with entry $E_{nm}^{t}$ and $\mathbf{D}_{t}$ is a $\mathcal{B}$-dimensional
vector with entry $D_{n}^{t}$. We can also introduce a regularization
parameter $\varsigma$ with a very small value as $\mathbf{E}_{t}+\varsigma\mathbf{I}_{\mathcal{B}}$.
This regularization parameter prevents $\mathbf{E}_{t}$ from becoming
a singular matrix which does not have an inverse.

Once we obtain the optimal action $\widetilde{a}_{t}^{*}$ from Equation
(\ref{eq:weight1}) and Equation (\ref{eq:basis1}), the optimal $Q$-function
value at time $t$ can be inferred by an expectation in terms of the
reward at time $t$ and the optimal $Q$-function value at time $t+1$
through the Bellman optimality equation (Equation (\ref{eq:Q bellman optimality}))
at the optimal action $\widetilde{a}_{t}^{*}$. However, in practice
we take the sample from our simulated path instead of the expectation
to infer the optimal $Q$-function value at time $t$ and this will
produce error $\mathsf{e}_{u}$ which is the deviation of the observed
value from the true value:

\begin{align}
Q_{t}^{\pi^{*}}\left(\widetilde{S}_{t},\widetilde{a}_{t}^{*}\right) & =\mathbb{E}_{t}\left[R_{t}\left(\widetilde{S}_{t},\widetilde{a}_{t}^{*},\widetilde{S}_{t+1}\right)+\gamma Q_{t+1}^{\pi^{\mathit{*}}}\left(\widetilde{S}_{t+1},\widetilde{a}_{t+1}^{*}\right)\right]\nonumber \\
 & =R_{t}\left(\widetilde{S}_{t}^{u},\widetilde{a}_{t}^{*},\widetilde{S}_{t+1}^{u}\right)+\gamma Q_{t+1}^{\pi^{\mathit{*}}}\left(\widetilde{S}_{t+1}^{u},\widetilde{a}_{t+1}^{*}\right)+\mathsf{e}_{u}.\label{eq:Q=00003DEsti+e}
\end{align}

The optimal $Q$-function value at time $t$ can be learned by reaching
the criteria of minimizing the sum of squared errors in all simulated
paths: 
\begin{align}
\varphi_{nt}^{*} & =\min_{\varphi_{nt}}\sum_{u=1}^{U}\mathsf{e}_{u}^{2}\nonumber \\
 & =\min_{\varphi_{nt}}\sum_{u=1}^{U}\left(F_{u}-\sum_{n=1}^{\mathcal{B}}\varphi_{nt}\psi_{n}\left(\widetilde{S}_{t}^{u}\right)\right)^{2},\label{square errors}
\end{align}
where $F_{u}=R_{t}\left(\widetilde{S}_{t}^{u},\widetilde{a}_{t}^{*},\widetilde{S}_{t+1}^{u}\right)+\gamma Q_{t+1}^{\pi^{\mathit{*}}}\left(\widetilde{S}_{t+1}^{u},\widetilde{a}_{t+1}^{*}\right)$.

The optimal coefficients $\varphi_{nt}^{*}$ is approximated by the
least square estimator $\widehat{\varphi}_{t}^{*}$:
\begin{align}
\varphi_{t}^{*} & \approx\widehat{\varphi}_{t}^{*}\nonumber \\
 & =\mathbf{G}_{t}^{-1}\mathbf{H}_{t},\label{eq:Least Square}
\end{align}

\begin{equation}
G_{nm}^{t}=\psi_{n}\left(\widetilde{S}_{t}^{u}\right)\psi_{m}\left(\widetilde{S}_{t}^{u}\right),\label{eq:G}
\end{equation}

\begin{equation}
H_{n}^{t}=\psi_{n}\left(\widetilde{S}_{t}^{u}\right)\left(R_{t}\left(\widetilde{S}_{t}^{u},\widetilde{a}_{t}^{*},\widetilde{S}_{t+1}^{u}\right)+\gamma Q_{t+1}^{\pi^{\mathit{*}}}\left(\widetilde{S}_{t+1}^{u},\widetilde{a}_{t+1}^{*}\right)\right),\label{eq:H}
\end{equation}
where $\varphi_{t}^{*}$ is a $\mathcal{B}$-dimensional vector with
entry $\varphi_{nt}^{*}$, $\mathbf{G}_{t}$ is a matrix of size $\mathcal{B}\times\mathcal{B}$
with entry $G_{nm}^{t}$ and $\mathbf{H}_{t}$ is a $\mathcal{B}$-dimensional
vector with entry $H_{n}^{t}$.

We are ultimately able to predict the option price $\mathcal{C}_{t}\left(\widetilde{S}_{t}\right)=-Q_{t}^{\pi^{\mathit{*}}}\left(\widetilde{S}_{t},\widetilde{a}_{t}^{*}\right)$
and the hedge position $\widetilde{a}_{t}^{*}$ backward recursively
at any time until the expiration date by triggering the following
algorithm:

\noindent\begin{minipage}[t]{1\columnwidth}%
\end{minipage}

\begin{tabular}{l}
\hline 
\textbf{\footnotesize{}Algorithm of learning option price and hedge
position}\tabularnewline
\hline 
\textbf{\footnotesize{}1: }{\footnotesize{}Calculate}\textbf{\footnotesize{}
$Q_{\mathsf{T}}^{\pi^{\mathit{*}}}\left(\widetilde{S}_{\mathsf{T}}^{u},\widetilde{a}_{\mathsf{T}}^{*}\right)$}{\footnotesize{}
with Eq. (\ref{eq:terminal condition}) and $\Pi_{\mathsf{T}}$ with
Eq. (\ref{eq:terminal portfolio value})}\textbf{\footnotesize{} }\tabularnewline
\textbf{\footnotesize{}2: for }{\footnotesize{}$t=\mathsf{T}-1$ }\textbf{\footnotesize{}to}{\footnotesize{}
$t=0$ }\textbf{\footnotesize{}do}\tabularnewline
\textbf{\footnotesize{}3: }{\footnotesize{}Calculate}\textbf{\footnotesize{}
$w_{t}^{*}$ }{\footnotesize{}with Eq.(\ref{eq:weight1}), $\widetilde{a}_{t}^{*}$
with Eq. (\ref{eq:basis1}) and $\Pi_{t}$ with Eq. (\ref{eq: portfolio values difference})}\tabularnewline
\textbf{\footnotesize{}4: }{\footnotesize{}Calculate}\textbf{\footnotesize{}
$R_{t}\left(\widetilde{S}_{t},\widetilde{a}_{t}^{*},\widetilde{S}_{t+1}\right)$
}{\footnotesize{}with Eq. (\ref{eq:Reward})}\textbf{\footnotesize{} }\tabularnewline
\textbf{\footnotesize{}5: }{\footnotesize{}Calculate}\textbf{\footnotesize{}
$\varphi_{t}^{*}$ }{\footnotesize{}with Eq.(\ref{eq:Least Square})
and $Q_{t}^{\pi^{\mathit{*}}}\left(\widetilde{S}_{t},\widetilde{a}_{t}^{*}\right)$
with Eq. (\ref{eq:basis2})}\tabularnewline
\textbf{\footnotesize{}6: end for}\tabularnewline
\hline 
\end{tabular}

\section{Experimental Evaluation}

We evaluate our entire algorithm/process on a task of hedging and
pricing an OTC European call option based on 100 shares of the underlying
GME stock with a period of 6 trading days and a strike price of 100
USD. Suppose that we have learned that the leading investors are the
same group of investors who took on the role as PI during Jan. 26,
2021-Feb. 02, 2021. Therefore, our first step is to probabilistically
imitate their previous successive decision making from the hidden
motives behind them through our visible-hidden Markov network as we
presented in the section \ref{sec:Inverse-PI's-decision}. The second
step is to transform likely evolutionary paths of PI's decision into
probable price paths of the underlying stock by our B-DNN inference-regression
process introduced in the section \ref{sec:Effect-of-the}. 
\begin{figure}
\subfloat[\label{fig:7a}]{\includegraphics[width=4.5cm,height=3.5cm]{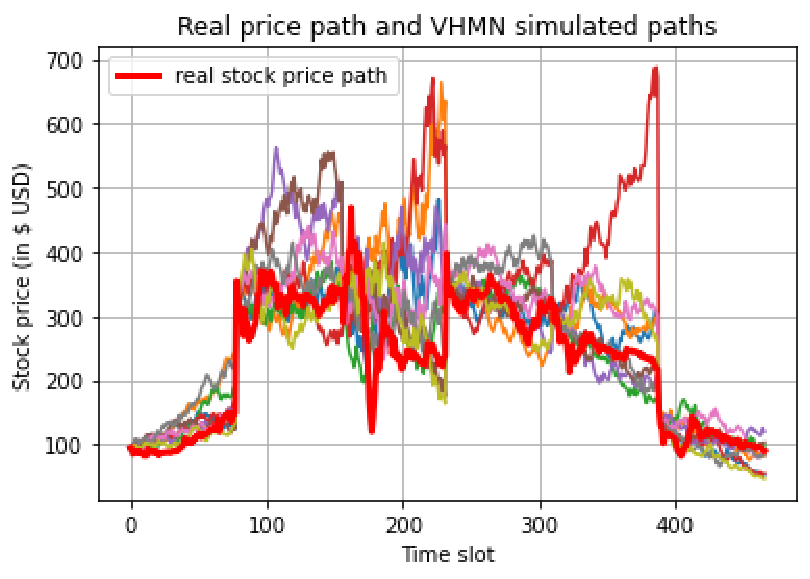}}\subfloat[\label{fig:7b}]{\includegraphics[width=4.5cm,height=3.5cm]{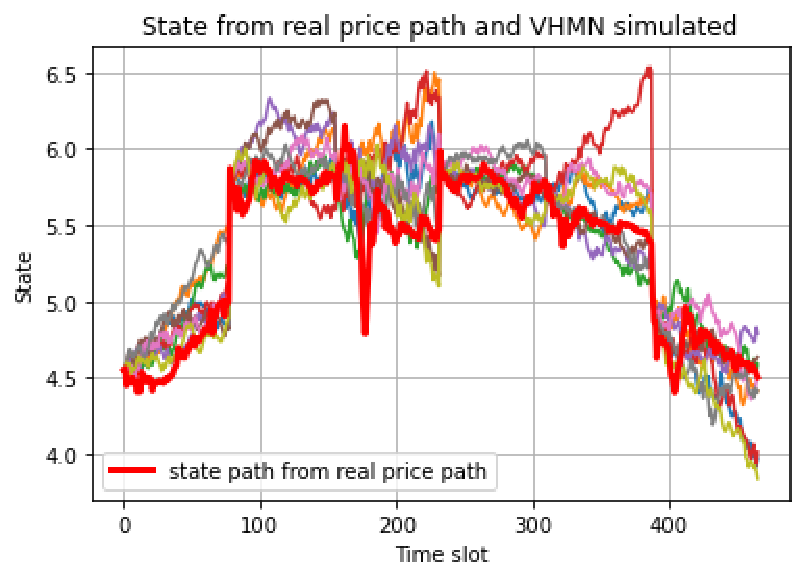}}\caption{Imitative GME price movements\label{fig:PricePathAll6days}}
\end{figure}
\begin{figure}
\subfloat[\label{fig:8a}]{\includegraphics[width=4.5cm,height=3.5cm]{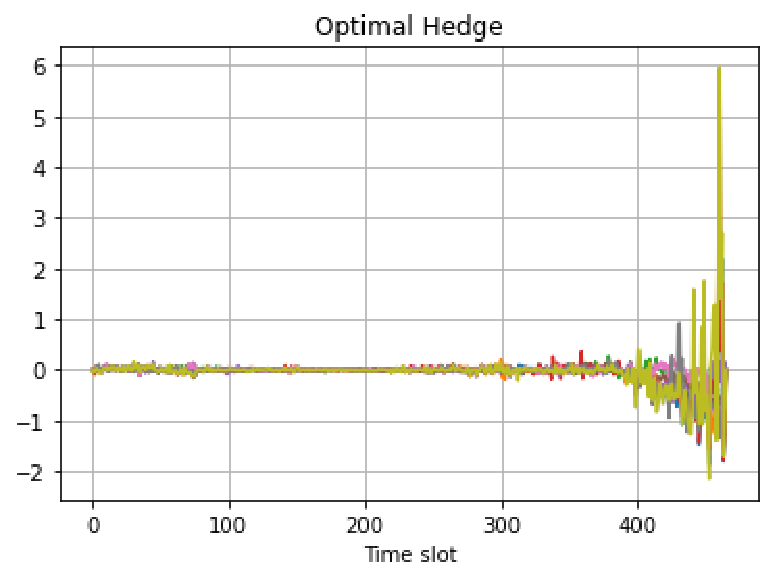}}\subfloat[\label{fig:8b}]{\includegraphics[width=4.5cm,height=3.5cm]{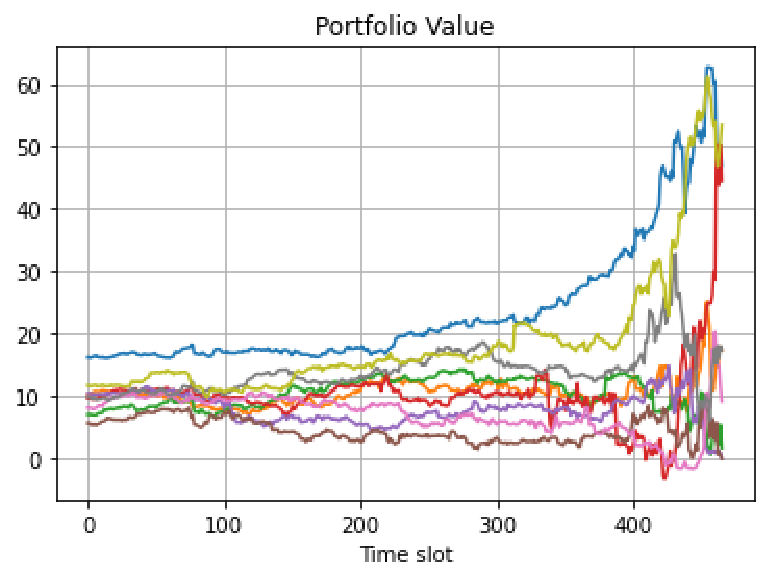}}\medskip{}

\subfloat[\label{fig:8c}]{\includegraphics[width=4.5cm,height=3.5cm]{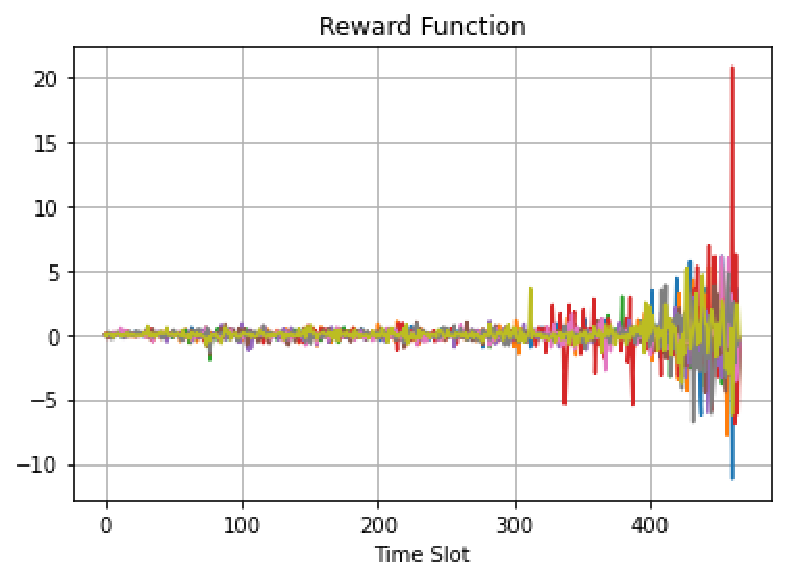}}\subfloat[\label{fig:8d}]{\includegraphics[width=4.5cm,height=3.5cm]{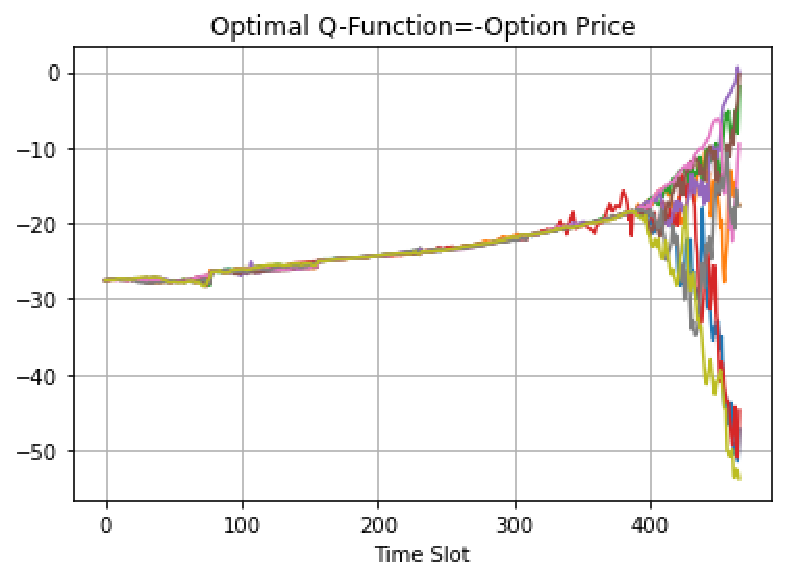}}

\caption{Reinforcement learn option price and hedging position ground upon
imitative PI's chronological decision making\label{fig:Pricing and hedging option }}
\end{figure}
For the above two steps, we train a B-DNN and a visible-hidden Markov
network on an episode basis with the training data set $\mathcal{\mathscr{D}}_{i,i=1\ldots6}$.
The simulated price paths along with the real price paths used as
the training data samples are compared in Figure \ref{fig:7a}. We
remove the constant drift from all the price paths and plot the real
state $\widetilde{S}_{t}$ and the simulated states $\widetilde{S}_{t}^{u}$
in Figure \ref{fig:7b}. The drift rate is calculated with $\mu=0.05$
and $\sigma_{s}=0.2926$. The value of $\sigma_{s}$ is derived from
averaging the daily VIX index from Jan. 26, 2021 to Feb. 02, 2021.
The total number of the simulated paths is 1000 and only 10 paths
are uniformly sampled to be shown here. The training samples are sampled
in each 5-minute time slot. The simulated paths are also with a time
resolution of 5 minutes. The price jumps that appear in Figure \ref{fig:7a}
and \ref{fig:7b} are from the pricing data in the first 5-min time
slot (i.e., 09:30 to 09:35). The open price and the volume in this
time slot deviate far from the average owing to their high dependence
on the pre-market trading and news. For this reason, the training
dataset does not include the pricing data in the first 5-min time
slot, and we interpolate the real value in the figures. The Visible-hidden
Markov network is abbreviated to VHMN in the figures for the benefit
of space. We can observe that the visible-hidden Markov network produces
imitative price paths that have a highly matched behavior pattern
with the real price path. As a consequence, the visible-hidden Markov
network is a powerful tool to cope with the intrinsic instability
in financial data by engendering imitative variations for each specific
scenario.

Our final phase of this task is to perform the algorithm of learning
option price and hedge position. We rebalance the portfolio every
5 minutes during 6 trading days. The risk-free interest rate $r=1.059\%$
is taken as the average of daily US 10-year bond yield from Jan. 26,
2021 to Feb. 02, 2021. We set $\kappa=0$ as we only hedge a mini
option based on 100 shares. We also give $\frac{1}{\eta}=0$ for having
pure risk hedge. $\varsigma=0.001$ is used to avoid the error of
inverting a singular matrix. The optimal hedging position is shown
in Figure (\ref{fig:8a}), where the value on the graph needs to be
multiplied by 100 to correspond to 100 shares. It is exhibited that
the option is more actively hedged when closer to the expiration date.
The value of the replicating portfolio jumps near the maturity date
and the hedging strategy effectively protects the portfolio against
the risk of running out of money as shown in Figure (\ref{fig:8b}).
The optimal $Q$-function value converges at -27.586 which means we
better sell the option at the price of 2758.6 USD as illustrated in
Figure (\ref{fig:8d}). 

\section{Conclusion}

In order to address the challenges posed by intrinsic low SNR and
instability in financial data, we innovatively exploit imitative cross-sectional
information to learn option price and hedging position with reinforcement
learning. When we turn to behavior finance, another challenge of identifying
the leading investor's behavior and the stock price change comes along.
To this end, we take advantage of the excellent features of B-DNN
to explore non-deterministic relations in a market data driven fashion.

\section*{Appendix A}

Denote likelihood times prior as $a\left(\boldsymbol{\theta}\right)\coloneqq p\left(\mathcal{\mathscr{D}}\mathrm{|\boldsymbol{\theta}}\right)p\left(\boldsymbol{\theta}\right)$
which satisfies 2nd-order sufficient conditions: $a^{\prime}\left(\boldsymbol{\theta}^{\ast}\right)=\boldsymbol{0}$
and $a^{\prime\prime}\left(\boldsymbol{\theta}^{\ast}\right)$ is
positive definite, where $\boldsymbol{\theta}^{\ast}$ is a strict
local maximizer. We do a Taylor series approximation of $\log a\left(\boldsymbol{\theta}\right)$
around the location of its maximum to give: 

\begin{equation}
\log a\left(\boldsymbol{\theta}\right)\approx\log a\left(\boldsymbol{\theta}^{\ast}\right)+\frac{1}{2}\left(\boldsymbol{\theta}-\boldsymbol{\theta}^{\ast}\right)^{T}\nabla_{\boldsymbol{\theta\theta}}^{2}\log a\left(\boldsymbol{\theta}^{\ast}\right)\left(\boldsymbol{\theta}-\boldsymbol{\theta}^{\ast}\right)\label{Taylor series approx}
\end{equation}

Plugging this truncated Taylor expansion of $\log a\left(\boldsymbol{\theta}\right)$
into the posterior $P\left(\mathrm{\boldsymbol{\theta}}|\mathcal{\mathscr{D}}\right)$
will demonstrate the posterior as the gaussian distribution $\mathcal{N}\left(\boldsymbol{\theta}\mid\boldsymbol{\theta}^{\ast},\left[-\nabla_{\boldsymbol{\theta\theta}}^{2}\log a\left(\boldsymbol{\theta}^{\ast}\right)\right]^{-1}\right)$:

\begin{align}
P\left(\mathrm{\boldsymbol{\theta}}|\mathcal{\mathscr{D}}\right) & =\frac{P\left(\mathcal{\mathscr{D}}\mathrm{|\boldsymbol{\theta}}\right)P\left(\boldsymbol{\theta}\right)}{\int P\left(\mathcal{\mathscr{D}}\mathrm{|\boldsymbol{\theta}}\right)P\left(\boldsymbol{\theta}\right)d\boldsymbol{\theta}}\nonumber \\
 & =\frac{\exp\left(\log a\left(\boldsymbol{\theta}\right)\right)}{\int\exp\left(\log a\left(\boldsymbol{\theta}\right)\right)d\boldsymbol{\theta}}\nonumber \\
 & \approx\frac{\exp\left(-\frac{1}{2}\left(\boldsymbol{\theta}-\boldsymbol{\theta}^{\ast}\right)^{T}\left[-\nabla_{\boldsymbol{\theta\theta}}^{2}\log a\left(\boldsymbol{\theta}^{\ast}\right)\right]\left(\boldsymbol{\theta}-\boldsymbol{\theta}^{\ast}\right)\right)}{\int\exp\left(-\frac{1}{2}\left(\boldsymbol{\theta}-\boldsymbol{\theta}^{\ast}\right)^{T}\left[-\nabla_{\boldsymbol{\theta\theta}}^{2}\log a\left(\boldsymbol{\theta}^{\ast}\right)\right]\left(\boldsymbol{\theta}-\boldsymbol{\theta}^{\ast}\right)\right)d\boldsymbol{\theta}}\nonumber \\
 & =\frac{\exp\left(-\frac{1}{2}\left(\boldsymbol{\theta}-\boldsymbol{\theta}^{\ast}\right)^{T}\left[-\nabla_{\boldsymbol{\theta\theta}}^{2}\log a\left(\boldsymbol{\theta}^{\ast}\right)\right]\left(\boldsymbol{\theta}-\boldsymbol{\theta}^{\ast}\right)\right)}{\sqrt{2\pi\left[-\nabla_{\boldsymbol{\theta\theta}}^{2}\log a\left(\boldsymbol{\theta}^{\ast}\right)\right]^{-1}}}\label{PosteriorGaussian}
\end{align}

When we enforce the likelihood and the prior belong to the exponential
family with the following expression $P\left(\mathscr{D}\mid\boldsymbol{\theta}\right)\coloneqq\frac{e^{-L\left(\boldsymbol{\theta},\mathscr{D}\right)}}{\int e^{-L\left(\boldsymbol{\theta},\mathscr{D}\right)}d\boldsymbol{\theta}}$
and $P\left(\boldsymbol{\theta}\right)\coloneqq\frac{e^{\mathit{-r}\left(\boldsymbol{\boldsymbol{\theta}}\right)}}{\int e^{\mathit{-r}\left(\boldsymbol{\boldsymbol{\theta}}\right)}d\boldsymbol{\theta}}$,
the gaussian distribution followed by the posterior becomes $\mathcal{N}\left(\boldsymbol{\theta}\mid\boldsymbol{\theta}^{\ast},\left[\nabla_{\boldsymbol{\theta\theta}}^{2}L\left(\boldsymbol{\theta}^{\ast},\mathscr{D}\right)+\lambda\mathbf{I}_{\mathit{M}}\right]^{-1}\right)$,
where $\mathbf{I}_{\mathit{M}}$ is an identity matrix of size $M\times M$.

\section*{Appendix B}

From Equation (\ref{eq:Bellman optimality equation}) and Equation
(\ref{eq:basis1}), we know $Q_{t}^{\pi}\left(\widetilde{S}_{t},\widetilde{a}_{t}\right)$
is a quadratic function of $\widetilde{a}_{t}$ and $\widetilde{a}_{t}$
is a linear function of $w_{nt}$. Therefore, finding the optimal
coefficients $w_{nt}^{*}$ which maximize $Q$-function value is equivalent
to solving the following equation $\frac{-\partial Q_{t}^{\pi}\left(\widetilde{S}_{t},\widetilde{a}_{t}\right)}{\partial w_{nt}}\mid_{w_{nt}=w_{nt}^{*}}=0$:

\begin{strip}

\begin{align}
\frac{-\partial Q_{t}^{\pi}\left(\widetilde{S}_{t},\widetilde{a}_{t}\right)}{\partial w_{nt}}\nonumber \\
=\frac{-\partial Q_{t}^{\pi}\left(\widetilde{S}_{t},\widetilde{a}_{t}\right)}{\partial\widetilde{a}_{t}}\frac{\partial\widetilde{a}_{t}}{\partial w_{nt}}\nonumber \\
=\mathbb{E}_{t}\left[\left(\bigtriangleup S_{t}+\kappa f_{\boldsymbol{\theta}^{*}}\left(F\right)S_{t}\right)\sum_{n=1}^{N}\psi_{n}\left(\widetilde{S}_{t}\right)\right]-2\eta\gamma\mathbb{E}_{t}\left[\left(\dot{\Pi}_{t+1}-\left(\widetilde{a}_{t}\bigtriangleup\dot{S}_{t}+\kappa f_{\boldsymbol{\theta}^{*}}\left(F\right)\widetilde{a}_{t}\dot{S}_{t}\right)\right)\left(-\bigtriangleup\dot{S}_{t}-\kappa f_{\boldsymbol{\theta}^{*}}\left(F\right)\dot{S}_{t}\right)\sum_{n=1}^{\mathcal{B}}\psi_{nt}\left(\widetilde{S}_{t}\right)\right]\nonumber \\
\approx\frac{1}{U}\sum_{u=1}^{U}\left[\left(\bigtriangleup S_{t}^{u}+\kappa f_{\boldsymbol{\theta}^{*}}\left(F\right)S_{t}^{u}\right)\sum_{n=1}^{\mathcal{B}}\psi_{nt}\left(\widetilde{S}_{t}^{u}\right)\right]\nonumber \\
\frac{-2\eta\gamma}{U}\sum_{u=1}^{U}\left[\left(\dot{\Pi}_{t+1}-\left(\widetilde{a}_{t}\bigtriangleup\dot{S}_{t}^{u}+\kappa f_{\boldsymbol{\theta}^{*}}\left(F\right)\widetilde{a}_{t}\dot{S}_{t}^{u}\right)\right)\left(-\bigtriangleup\dot{S}_{t}^{u}-\kappa f_{\boldsymbol{\theta}^{*}}\left(F\right)\dot{S}_{t}^{u}\right)\sum_{n=1}^{\mathcal{B}}\psi_{nt}\left(\widetilde{S}_{t}^{u}\right)\right] & ,\label{eq:dQ/dcoefficients}
\end{align}

\begin{align}
\frac{-\partial Q_{t}^{\pi}\left(\widetilde{S}_{t},\widetilde{a}_{t}\right)}{\partial w_{nt}}\mid_{w_{nt}=w_{nt}^{*}}=0 & \Rightarrow\nonumber \\
\sum_{u=1}^{U}\left[\left(\frac{\bigtriangleup S_{t}^{u}}{2\eta\gamma}+\dot{\Pi}_{t+1}\varXi+\frac{\kappa f_{\boldsymbol{\theta}^{*}}\left(F\right)S_{t}^{u}}{2\eta\gamma}\right)\sum_{n=1}^{\mathcal{B}}\psi_{n}\left(\widetilde{S}_{t}^{u}\right)\right]=\nonumber \\
\sum_{u=1}^{U}\left[\left(\sum_{m=1}^{\mathcal{B}}w_{mt}^{*}\psi_{m}\left(\widetilde{S}_{t}^{u}\right)\bigtriangleup\dot{S}_{t}\varXi+\kappa f_{\boldsymbol{\theta}^{*}}\left(F\right)\sum_{m=1}^{\mathcal{B}}w_{mt}^{*}\psi_{m}\left(\widetilde{S}_{t}^{u}\right)\dot{S}_{t}\varXi\right)\sum_{n=1}^{\mathcal{B}}\psi_{n}\left(\widetilde{S}_{t}^{u}\right)\right] & \Rightarrow\nonumber \\
D_{n}^{t}=\sum_{m=1}^{\mathcal{B}}E_{nm}^{t}w_{mt}^{*},n=1\ldots\mathcal{B} & \Rightarrow\nonumber \\
w_{t}^{*}=\mathbf{E}_{t}^{-1}\mathbf{D}_{t} & ,\label{eq:DEw}
\end{align}
where $w_{t}^{*}$ is a $\mathcal{B}$-dimensional vector with entry
$w_{nt}^{*}$. $\mathbf{E}_{t}$ and $\mathbf{D}_{t}$ are respectively
a matrix of size $\mathcal{B}\times\mathcal{B}$ and a $\mathcal{B}$-dimensional
vector. Their entries $E_{nm}^{t}$ and $D_{n}^{t}$ are formulated
below:

\[
E_{nm}^{t}=\sum_{u=1}^{U}\left[\left(\psi_{m}\left(\widetilde{S}_{t}^{u}\right)\bigtriangleup\dot{S}_{t}\varXi+\kappa f_{\boldsymbol{\theta}^{*}}\left(F\right)\psi_{m}\left(\widetilde{S}_{t}^{u}\right)\dot{S}_{t}\varXi\right)\sum_{n=1}^{\mathcal{B}}\psi_{n}\left(\widetilde{S}_{t}^{u}\right)\right],
\]

\[
D_{n}^{t}=\sum_{u=1}^{U}\left[\left(\frac{\bigtriangleup S_{t}^{u}}{2\eta\gamma}+\dot{\Pi}_{t+1}\varXi+\frac{\kappa f_{\boldsymbol{\theta}^{*}}\left(F\right)S_{t}^{u}}{2\eta\gamma}\right)\sum_{n=1}^{\mathcal{B}}\psi_{n}\left(\widetilde{S}_{t}^{u}\right)\right],
\]

\[
\varXi=\left(\bigtriangleup\dot{S}_{t}+\kappa f_{\boldsymbol{\theta}^{*}}\left(F\right)\dot{S}_{t}\right).
\]
\end{strip}

\end{document}